\documentclass[11pt,a4paper,reqno]{amsart}

\usepackage[mathscr]{euscript}
 \usepackage[cal=boondoxo]{mathalfa}
\usepackage{amssymb}
\usepackage{amsthm}
\usepackage{amsxtra}
\usepackage{bbm}
\usepackage{mathtools}
\usepackage{stmaryrd}

\usepackage{extarrows}
\usepackage{enumitem}
\usepackage{tikz-cd}
\usepackage{url}
\usepackage{booktabs}
\usepackage{graphicx}
\usepackage{resizegather}
\usepackage{arydshln}
\usepackage{floatrow}
\usepackage{soul}
\usepackage{relsize,etoolbox}

\numberwithin{equation}{section}
\newcommand\scalemath[2]{\scalebox{#1}{\mbox{\ensuremath{\displaystyle #2}}}}

\usepackage[pdftex,
                paper=a4paper,
                portrait=true,
                textwidth=165mm,
                textheight=235mm,
                tmargin=2.5cm,
                marginratio=1:1]{geometry}
                
\theoremstyle{definition}

\theoremstyle{remark}

\newtheorem*{acknowledgements}{Acknowledgements}

\newcommand{\CC}{\ensuremath{\mathbb{C}}}

\newcommand{\ZZ}{\ensuremath{\mathbb{Z}}}

\renewcommand{\Re}{\operatorname{Re}}
\newcommand{\uS}{\operatorname{S}}
\newcommand{\SL}{\operatorname{SL}}
\newcommand{\SO}{\operatorname{SO}}
\newcommand{\SU}{\operatorname{SU}}
\newcommand{\U}{\operatorname{U}}
\newcommand{\tr}{\operatorname{Tr}}
\newcommand{\ord}{\operatorname{ord}}

\DeclareMathAlphabet{\mathbfit}{OML}{cmm}{b}{it}
\newcommand*{\bigcdot}{\raisebox{-0.25ex}{\scalebox{1.3}{\,$\cdot$\,}}}

\title{A $\U(2) \times \U(3)$ gauge theory extension of the standard model}

\author{Rafael Herrera
and Alexander Quintero V\'{e}lez} 
  
\address{Rafael Herrera, Centro de Investigaci\'{o}n en Matem\'{a}ticas \\
A. P. 402, 36000 Guanajuato\\
Guanajuato, M\'{e}xico}
 \email{rherrera@cimat.mx}
 
\address{Alexander Quintero V\'{e}lez, Escuela de Matem\'{a}ticas\\
Universidad Nacional de Colombia Sede Medell\'{i}n\\
Calle 59a \# 63-20\\
Medell\'{i}n, Colombia}
 \email{aquinte2@unal.edu.co}

\begin{document}

\begin{abstract}
We consider an extension of the standard model based on the group $\U(2) \times \U(3)$, which is naturally compatible with the standard model interacting-particle representations and the spontaneous symmetry breaking of $\U(2) \times \U(3)$ to an electrostrong $\U(3)$. In its minimal version, the model only adds one extra $\U(1)$ gauge boson and it implies that the hypercharge is distributed between the factors of the hyperweak and hyperstrong forces. We show that the anomaly cancellation condition can be solved by adding exotic fermions associated with a $16$-dimensional representation of $\U(2) \times \U(3)$. A brief discussion of the mechanism of the spontaneous breakdown of $\U(2) \times \U(3)$ in the gauge boson sector is given.  
\end{abstract}

 \maketitle

\section{Introduction}\label{intro}
The standard model of electroweak and strong interactions has been successful in explaining essentially all known experimental facts with a minimal number of particles.  In spite of its great success, the model is not regarded as a final theory, and many generalisations have been proposed in the last four decades in order to accommodate new experimental results and also for aesthetic reasons (see, for example, \cite{Langacker09}).

In this paper we present an extension of the standard model based on the symmetry group $\U(2) \times \U(3)$. We were led to this symmetry group by the following considerations. We first noted that, by the principle of confinement, all observable states must be ``white'', which means that they must be acted on trivially by $\SU(3)$ (see the discussion on page 496 of~\cite{BaezHuerta10}). This implied  that we should rewrite the standard model interacting-particle representations (see Tables~\ref{tab:1} and~\ref{tab:2}). Using well known properties of the exterior powers of representations, we then observed that by ``distributing'' the hypercharge judiciously between the weak and strong factors, we arrived at a set of representations which displays great regularity (see Table~\ref{tab:3}). In particular, we found that the representation one should use to describe a single generation of fermions in the standard model is given by the exterior algebra of a representation of $\U(1) \times \SU(2)$ tensored by the odd part of the exterior algebra of a representation of $\U(1) \times \SU(3)$ (see Section~\ref{Sect:2.1} below for details). At this point, it became clear that we were really dealing with representations of $\U(2) \times \U(3)$ restricted to a suitable subgroup, hereafter denoted by $\uS(\U(2) \times \U(3))$ (see \cite{BaezHuerta10}  and  \cite{RV00}). Thus, we sought to build a $\U(2) \times \U(3)$ theory by mapping the standard model group $\U(1) \times \SU(2) \times \SU(3)$ into $\U(2) \times \U(3)$ as follows:
$$
(e^{i\theta}, A, B) \longmapsto (e^{i3\theta}A, e^{-i2\theta}B).
$$
In this way, we make sure that the aforementioned subgroup $\uS(\U(2) \times \U(3))$ is isomorphic to the image of $\U(1) \times \SU(2) \times \SU(3)$. 

The particle content of the $\U(2) \times \U(3)$ model we consider includes the fundamental fermions of the standard model from a $16$-dimensional representation of $\U(2) \times \U(3)$, and exotic fermions from a different $16$-dimensional representation of $\U(2) \times \U(3)$ with a particular choice of chiralities. These exotic fermions are added in order to ensure that the model is free from chiral anomalies as well as to ensure that it is as symmetric as possible. The space of exotic fermions chosen in this paper is only a $16$-dimensional subspace of the much bigger $512$-dimensional representation (see Section~\ref{Sect:2.3}) suggested by our model and is, of course, not unique. Many other choices of exotic fermions are possible which may fulfill more phenomenological constraints. As a consistency check, we show that the given $\U(2) \times \U(3)$ representations of both the fundamental and exotic fermions are well behaved when $\U(2) \times \U(3)$ is spontaneously broken to an electrostrong $\U(3)$. 

In addition, there will be thirteen gauge bosons, twelve of which will correspond to the ordinary gauge bosons of the standard model after symmetry breaking, and one more corresponding to an extra $\U(1)$ gauge boson. The symmetry breaking is implemented by introducing two self-interacting Higgs fields, a ``singlet'' and a ``doublet'' in physics terminology. A geometric description of these Higgs fields as cross sections of some associated vector bundles will be given. Within such a framework, the symmetry breaking will be characterised in terms of a reduction of a $\U(2) \times \U(3)$ principal bundle, on which the  model is set up, to appropriate subbundles. These subbundles are defined in terms of the given Higgs fields and their corresponding self-interaction potentials. Perhaps, it should be remarked that in this analysis we shall be only dealing with the interactions between the gauge fields and the Higgs fields, while ignoring the fermionic multiplets. The symmetry breaking in the fermionic sector will be studied in detail in a forthcoming paper. 

We would like to conclude the introduction by emphasising that, during the development of the theory, we took the standard model as our starting point, did not add any object to the model that the gauge symmetry group itself did not suggest, and pursued the greatest simplicity possible. Moreover, we think that we have addressed, in a mathematical sense, some of the questions raised by Baez and Huerta in the last paragraph of Section~3 of \cite{BaezHuerta10}:
\begin{quote}
``The representation
of $G_{\mathrm{SM}}$ used in the Standard Model seems ad hoc.\footnote{Here $G_{\mathrm{SM}}$ refers to the standard model group.} Why this one? Why are all those seemingly arbitrary hypercharges floating around, mucking up some otherwise simple representations? Why do both leptons and quarks come in left- and right-handed varieties, which transform so differently? Why do quarks come in charges which are in units $\frac{1}{3}$ times an electron's charge? Why are there the same number of quarks and leptons?''
\end{quote}
Indeed, as previously described, the distribution of the hypercharge and the properties of exterior powers do provide regularity to the multiplet representations in such a way that:  
\begin{itemize}
\item the hypercharges of the fundamental fermions no longer look arbitrary;
\item the interacting-particle representation for a single generation of fundamental fermions, taking no account of chirality, factors into a tensor product of an exterior algebra and the odd part of another exterior algebra;
\item  the numbers $2$, $3$ and $6$, or alternatively $\frac{1}{2}$, $\frac{1}{3}$ and $\frac{1}{6}$, that appear everywhere in the
calculations of the model actually depend on the $2$-fold, $3$-fold and $6$-fold covering group homomorphisms intrinsically present 
in the standard model group and its representation content, which are expressed succinctly in the defining map we are using to build the model.
\end{itemize}
Let us point out that in their expository paper, Baez and Huerta explain the well known answers to the questions provided by the grand unified $\SU(5)$ model~\cite{GeorgiGlashow74} and the grand unified $\SO(10)$ model~\cite{Georgi75,FritzschMinkowski75}. Our answers differ from those obtained by grand unified theories since, in our model, the gauge group $\U(2) \times \U(3)$ has two factors, so is not a real unification, as it requires four independent coupling constants. We are then left with new questions regarding the new exotic particles introduced for anomaly cancellations, as well as the phenomenological implications of the model. We shall address them in the near future.

The paper is organised as follows. In Section~\ref{Sect:1} we collect the notation and preliminaries used in later sections. In Section~\ref{Sect:2} we present the model, i.e., the classification of the fundamental and exotic fermions into left- and right-handed multiplets. We shall also describe the hypercharge assignments of the fermionic multiplets and the spontaneous symmetry breaking of $\U(2) \times \U(3)$ to the electrostrong $\U(3)$. Section~\ref{Sect:3} is devoted to the geometric description of the model, the study of the mechanism of the spontaneous breakdown of $\U(2) \times \U(3)$ and the gauge boson mass generation. The summary and conclusions are given in Section~\ref{Sect:4}.

{\small
\begin{acknowledgements}
The first named author is partially supported by grants from the Mexican National Council of Science and Technology (CONACyT) and the Italian National Institute of Nuclear Physics (INFN). He would like to thank the International Centre for Theoretical Physics, the Institut des Hautes \'Etudes Scientifiques, the Scuola Internazionale Superiore di Studi Avanzati for their hospitality and support. The authors would also like to thank Diego Restrepo Quintero for his very valuable insights.
\end{acknowledgements}
}

\section{Preliminaries}\label{Sect:1}

In this section we present the notation and some preliminary material which will be used throughout the paper. For further details, we refer to \cite{AGVM11,BaezHuerta10, BtD03, Ticciati99, McCabe11}.

\subsection{Notation from representation theory}\label{Sect:1.1}
As is well known, each of the irreducible representations of the gauge group of the standard model, $\U(1)\times \SU(2) \times \SU(3)$, is a tensor product of irreducible representations of the individual factors. Thus, it will be convenient to introduce the following notation.

\begin{itemize}[leftmargin=20pt]
\item The standard representation $\lambda$ of $\U(1)$ on $\CC^1$ maps $e^{i\theta}$ to $e^{i\theta}$, acting on $\CC^1$ by complex multiplication. For any positive integer $n$, one can form the $n$-fold tensor product $\lambda^n$ of $\lambda$, in which $e^{i\theta}$ acts as $e^{ i n \theta}$ upon $\CC^1$. The pertinent representation space will be denoted by $\CC^1_n$. In the event that $n$ is a negative integer, one can form the $n$-fold tensor product $\lambda^{-n}$ of the conjugate representation $\overline{\lambda}$, in which $e^{i\theta}$ acts as $e^{- i n \theta}$ upon $\CC^1$. Its representation space will be denoted by $\CC^1_{-n}$. Note that $\lambda^0$ is the trivial representation of $\U(1)$ which we shall denote by $1_{\U(1)}$.

\item The standard representation of $\SU(2)$ on $\CC^2$ will be denoted by $\tau$. It induces a representation $\bigwedge^2 \tau$ of $\SU(2)$ on the two-fold exterior power $\bigwedge^2 \CC^2$, which provides the same representation as the trivial representation of $\SU(2)$. We shall denote the latter by $1_{\SU(2)}$.

\item The standard representation of $\SU(3)$ on $\CC^3$ will be denoted by $\rho$. Such a representation induces a representation $\bigwedge^2 \rho$ of $\SU(3)$ on the two-fold exterior power $\bigwedge^2 \CC^3$. It also induces a representation $\bigwedge^3 \rho$ of $\SU(3)$ on the three-fold exterior power $\bigwedge^3 \CC^3$, which provides the same representation as the trivial representation of $\SU(3)$. We shall denote the latter by $1_{\SU(3)}$.
\end{itemize}

In order to account for the distinct handedness (``chirality'') of interacting elementary particles, these representations must be supplemented with the two inequivalent, irreducible representations of $\SL(2,\CC)$ on $\CC^2$, which we shall denote by $\sigma_{\mathnormal{L}}$ and $\sigma_{\mathnormal{R}}$, and which are known as the left- and right-handed spinor representations of $\SL(2,\CC)$. For easy reference we recall that $\sigma_{\mathnormal{L}}$ is just the standard representation of $\SL(2,\CC)$ on $\CC^2$ while $\sigma_{\mathnormal{R}}$ is the representation of $\SL(2,\CC)$ on $\CC^2$ that sends each $A \in \SL(2,\CC)$ to the inverse of its conjugate transpose $A^{*-1}$. We shall also consider the direct sum of the spinor representations $\sigma=\sigma_{\mathnormal{L}} \oplus \sigma_{\mathnormal{R}}$ of $\SL(2,\CC)$, which is often known as the Dirac spinor representation of $\SL(2,\CC)$.

\subsection{The standard model interacting-particle representations}
In the standard model of electromagnetic, weak, and strong interactions, a select collection of finite-dimensional irreducible representations of $\U(1) \times \SU(2) \times \SU(3)$ define the set of interacting elementary particles. These irreducible representations are said to define the elementary particle multiplets. Using the notation defined above, the finite-dimensional irreducible representations of $\U(1) \times \SU(2) \times \SU(3)$ which are conventionally associated with the particles in the first fermion generation are described in Table~\ref{tab:1}, where subscripts $L$ and $R$ refer to left- and right-handed chiralities respectively (compare with the table in \cite[\S 2.4]{BaezHuerta10}).
\begin{table}[!htbp]
\begin{tabular}{lcc}
\hline
\hline
Name                     & Symbol                 & Representation                              
	 \\
	 \hline 
	 \\
         Left-handed leptons      & $\left(\begin{array}{c} \nu_{\mathnormal{L}} \\ e_{\mathnormal{L}} \end{array}\right)$                 & $\lambda^{-3} \otimes \tau \otimes 1_{\SU(3)}$ \\
         \\                                                               
         Left-handed quarks       & $\left(\begin{array}{c} u_{\mathnormal{L}} \\ d_{\mathnormal{L}} \end{array}\right)$      & $\lambda \otimes \tau \otimes \rho$ \\
         \\                                                               
         Right-handed neutrino    & $\nu_{\mathnormal{R}}$                & $1_{\U(1)} \otimes 1_{\SU(2)} \otimes 1_{\SU(3)}$ \\
	 \\                                                               
         Right-handed electron    & $e_{\mathnormal{R}}$                & $\lambda^{-6} \otimes 1_{\SU(2)} \otimes 1_{\SU(3)}$ \\
	 \\                                                               
         Right-handed up quark   & $u_{\mathnormal{R}}$  &  $\lambda^{4} \otimes 1_{\SU(2)} \otimes \rho$ \\
	 \\                                                               
         Right-handed down quark & $d_{\mathnormal{R}}$  & $\lambda^{-2} \otimes 1_{\SU(2)} \otimes \rho$ \\ \\
	 \hline
	\end{tabular}
	\vspace{-5pt}
\caption{Fundamental fermions as representations of
$\U(1) \times \SU(2) \times \SU(3)$} \label{tab:1}
\end{table}

As the notation suggests, each of these representations is tensored with an irreducible, finite-dimensional representation of $\SL(2,\CC)$. In the case of the left-handed multiplets, the representation of the gauge group is tensored with the left-handed spinor representation $\sigma_{\mathnormal{L}}$ of $\SL(2,\CC)$, and in the case of the right-handed multiplets, the representation of the gauge group is tensored with the right-handed spinor representation $\sigma_{\mathnormal{R}}$ of $\SL(2,\CC)$. If we take the direct sum of all the irreducible representations, we obtain the following interacting-particle representation for the first fermion generation:
\begin{align*}
&\left( \lambda^{-3}\otimes \tau\otimes 1_{\SU(3)} \otimes \sigma_{\mathnormal{L}}\right)\oplus \left( \lambda\otimes \tau\otimes \rho \otimes \sigma_{\mathnormal{L}}\right) \oplus \left(1_{\U(1)}\otimes 1_{\SU(2)} \otimes 1_{\SU(3)}  \otimes \sigma_{\mathnormal{R}}\right) \\
&\quad \oplus \left( \lambda^{-6}\otimes 1_{\SU(2)}\otimes 1_{\SU(3)} \otimes \sigma_{\mathnormal{R}}\right) \oplus  \left( \lambda^{4}\otimes 1_{\SU(2)}\otimes \rho \otimes \sigma_{\mathnormal{R}}\right) \oplus  \left( \lambda^{-2}\otimes 1_{\SU(2)}\otimes \rho \otimes \sigma_{\mathnormal{R}}\right).
\end{align*}

\section{The $\U(2) \times \U(3)$ model}\label{Sect:2}
In this section, we present a gauge-theory model for leptons and quarks based on the group $\U(2) \times \U(3)$. All fermions are assigned to tensor products of exterior powers of the standard representations of the individual factors. The classification of the leptons and quarks into left- and right-handed multiplets (with a right-left asymmetry) is compatible with that of the standard model of electromagnetic, weak and strong interactions based on $\U(1) \times \SU(2) \times \SU(3)$. We also show that the gauge symmetry can be broken to that in which only electromagnetic and strong forces are ``turned on''.

\subsection{Heuristics}\label{Sect:2.1}
The model's point of departure is the observation that, by the principle of confinement, all observable states must be ``white'' with respect to the $\SU(3)$ ``color charge'', i.e., invariant under the action of $\SU(3)$. For us, this means the need to distinguish between the two trivial representations $1_{\SU(3)}$ and $\bigwedge^3 \rho$ of $\SU(3)$. Motivated by this consideration, we shall also distinguish the two trivial representations $1_{\SU(2)}$ and $\bigwedge^2 \tau$ of $\SU(2)$. Thus, we propose that the fundamental fermions are partitioned into multiplets by the finite-dimensional irreducible representations of $\U(1) \times \SU(2) \times \SU(3)$ described in Table \ref{tab:2}.
\begin{table}[!htbp]
\begin{tabular}{lcc}
\hline
\hline
Name                     & Symbol                 & Representation                              
	 \\
	 \hline 
	 \\
         Left-handed leptons      & $\left(\begin{array}{c} \nu_{\mathnormal{L}} \\ e_{\mathnormal{L}} \end{array}\right)$                 & $\lambda^{-3} \otimes \tau \otimes \boldsymbol{\bigwedge^3\rho}$ \\
         \\                                                               
         Left-handed quarks       & $\left(\begin{array}{c} u_{\mathnormal{L}} \\ d_{\mathnormal{L}} \end{array}\right)$      & $\lambda \otimes \tau \otimes \rho$ \\
         \\                                                               
         Right-handed neutrino    & $\nu_{\mathnormal{R}}$                & $1_{\U(1)} \otimes \boldsymbol{\bigwedge^2\tau} \otimes 
         \boldsymbol{\bigwedge^3\rho}$ \\
	 \\                                                               
         Right-handed electron    & $e_{\mathnormal{R}}$                & $\lambda^{-6} \otimes 1_{\SU(2)} \otimes \boldsymbol{\bigwedge^3\rho}$ \\
	 \\                                                               
         Right-handed up quark   & $u_{\mathnormal{R}}$  &  $\lambda^{4} \otimes \boldsymbol{\bigwedge^2\tau} \otimes \rho$ \\
	 \\                                                               
         Right-handed down quark & $d_{\mathnormal{R}}$  & $\lambda^{-2} \otimes 1_{\SU(2)} \otimes \rho$ \\ \\
	 \hline
	\end{tabular}
	\vspace{-5pt}
\caption{Fundamental fermions as representations of
$\U(1) \times \SU(2) \times \SU(3)$} \label{tab:2}
\end{table}

By postulating such a distinction between trivial representations, we can manipulate the representations in Table \ref{tab:2} for the first fermion generation as follows. Let us momentarily forget about tensoring with the left- and right-handed spinor representations of $\SL(2,\CC)$, and consider the direct sum of these irreducible representations:
 \begin{align*}
&\left( \lambda^{-3} \otimes \tau \otimes \textstyle\bigwedge^3\rho \right) \oplus\left( \lambda \otimes \tau \otimes \rho \right) \oplus \left( \textstyle 1_{\U(1)} \otimes \bigwedge^2\tau \otimes \bigwedge^3\rho \right) \\
 &\quad\oplus \left(\textstyle\lambda^{-6} \otimes 1_{\SU(2)} \otimes \bigwedge^3\rho \right) \oplus \left(\textstyle\lambda^{4} \otimes \bigwedge^2\tau \otimes \rho\right) \oplus \left(\lambda^{-2} \otimes 1_{\SU(2)} \otimes \rho \right).
\end{align*}
After some rearrangement, this direct sum may be written as
\begin{align*}
&\left[ \textstyle\bigwedge^0\left( \lambda^3 \otimes \tau \right) \oplus \bigwedge^1\left( \lambda^3 \otimes \tau \right) \oplus \bigwedge^2\left( \lambda^3 \otimes \tau \right)\right] \otimes \left[ \textstyle\bigwedge^1\left( \lambda^{-2} \otimes \rho \right) \oplus \bigwedge^3\left( \lambda^{-2} \otimes \rho \right) \right] \\
&\qquad= \textstyle\bigwedge^{\bullet}\left(\lambda^3 \otimes \tau\right) \otimes \bigwedge^{\mathrm{odd}}\left( \lambda^{-2} \otimes \rho \right).
\end{align*}
Note the regularity and uniformity that has been achieved. Thus, the previous table should read, instead, as described in Table \ref{tab:3}.
\begin{table}[!htbp]
\begin{tabular}{lcc}
\hline
\hline
Name                     & Symbol                 & Representation                              
	 \\
	 \hline 
	 \\
         Left-handed leptons      & $\left(\begin{array}{c} \nu_{\mathnormal{L}} \\ e_{\mathnormal{L}} \end{array}\right)$                 & $\bigwedge^1\left( \lambda^{3} \otimes \tau \right)\otimes \bigwedge^3\left( \lambda^{-2} \otimes \rho \right)$ \\
         \\                                                               
         Left-handed quarks       & $\left(\begin{array}{c} u_{\mathnormal{L}} \\ d_{\mathnormal{L}} \end{array}\right)$      & $\bigwedge^1\left( \lambda^{3} \otimes \tau \right)\otimes \bigwedge^1\left( \lambda^{-2} \otimes \rho \right)$ \\
         \\                                                               
         Right-handed neutrino    & $\nu_{\mathnormal{R}}$                & $\bigwedge^2\left( \lambda^{3} \otimes \tau \right)\otimes \bigwedge^3\left( \lambda^{-2} \otimes \rho \right)$ \\
	 \\                                                               
         Right-handed electron    & $e_{\mathnormal{R}}$                & $\bigwedge^0\left( \lambda^{3} \otimes \tau \right)\otimes \bigwedge^3\left( \lambda^{-2} \otimes \rho \right)$ \\
	 \\                                                               
         Right-handed up quark   & $u_{\mathnormal{R}}$  &  $\bigwedge^2\left( \lambda^{3} \otimes \tau \right)\otimes \bigwedge^1\left( \lambda^{-2} \otimes \rho \right)$ \\
	 \\                                                               
         Right-handed down quark & $d_{\mathnormal{R}}$  & $\bigwedge^0\left( \lambda^{3} \otimes \tau \right)\otimes \bigwedge^1\left( \lambda^{-2} \otimes \rho \right)$ \\ \\
	 \hline
	\end{tabular}
	\vspace{-5pt}
\caption{Fundamental fermions as representations of
$\U(1) \times \SU(2) \times \SU(3)$} \label{tab:3}
\end{table}

It is interesting to observe that, in the representations of Table \ref{tab:3} to which the fundamental fermions are assigned, the hypercharge gets split into a ``weak doublet'' and a ``colored triplet'' part. This will be a crucial feature in our considerations below.

\subsection{The gauge group $\U(2) \times \U(3)$}\label{Sect:2.2}
As previously mentioned, the gauge group for our model is $\U(2) \times \U(3)$. This group contains two independent $\U(1)$ factors  which may be taken as
\begin{align*}
\U(1)_{\mathnormal{L}} &= \left\{ \left(\begin{array}{cc} e^{i\alpha} & 0 \\ 0 & e^{i\alpha}   \end{array} \right) \, \Bigg\vert\, \alpha \in [0,2\pi)\right\} \subset \U(2),\\
\U(1)_{\mathnormal{C}}&=\left\{ \left(\begin{array}{ccc} e^{i\beta} & 0 & 0 \\ 0 & e^{i\beta} & 0 \\ 0 & 0 & e^{i\beta}  \end{array} \right) \, \Bigg\vert\, \beta \in [0,2\pi)\right\} \subset \U(3),
\end{align*}
where the subscripts $L$ and $C$ stand for ``left-handed chirality'' and ``color'', respectively, and are there to remind us that the group $\U(2)$ describes a ``hyperweak'' force, while the group $\U(3)$ describes a ``hyperstrong'' force. In order to yield a ``realistic'' theory of leptons and quarks, the $\U(2) \times \U(3)$ group must contain the gauge group of the standard model as a subgroup. Whilst it is often said that the latter group is $\U(1) \times \SU(2) \times \SU(3)$, the strict standard model group is actually a quotient $(\U(1) \times \SU(2) \times \SU(3))/\ZZ_6$ with respect to a finite central $\ZZ_6$ subgroup. In our current setting, this quotient group has a neat description as the subgroup $\uS(\U(2) \times \U(3))$ of $\U(2) \times \U(3)$ consisting of pairs $(u,v) \in \U(2) \times \U(3)$ such that
$$
(\det u)(\det v)=1.
$$
In such a description, the hypercharge group $\U(1)_{\mathnormal{Y}}$ is taken to be
$$
\U(1)_{\mathnormal{Y}} = \big\{ \big(e^{i3\theta} I_2, e^{-i2\theta} I_3\big) \, \big\vert \, \theta \in [0,2 \pi) \big\},
$$
where $I_2$ is de $2 \times 2$ identity matrix and $I_3$ is the $3 \times 3$ identity matrix. Compare this with the definition in \cite[\S 2]{RV00}.

On the other hand, we must also incorporate the fact that the hyperweak force has to undergo spontaneous symmetry breaking. The unbroken symmetry group is an ``electrostrong'' $\U(3)$ realised as the following subgroup of $\uS(\U(2) \times \U(3))$:
$$
\U(3)_{\mathnormal{Q}} = \left\{ \left(\left(\begin{array}{cc} (\det v)^{2} & 0 \\ 0 & 1   \end{array} \right), (\det v)^{-1}v \right) \, \Bigg\vert\, v \in \U(3) \right\}.
$$
Within this framework, the electromagnetic group $\U(1)_{\mathnormal{Q}}$  is taken to be
$$
\U(1)_{\mathnormal{Q}} = \left\{ \left( \left(\begin{array}{cc} e^{i 6 \theta} & 0 \\ 0 & 1   \end{array} \right), e^{-i 2\theta} I_3 \right)  \, \Bigg\vert\,  \theta \in [0,2\pi) \right\}.
$$
Compare again with \cite[\S 2]{RV00}. Later, we shall see that this choice brings about the correct charges to the fundamental fermions of the model.

Next we take up the question as to which specific realization of the group $\U(2) \times \U(3)$ we should use. To that end, consider the homomorphism $\Phi$ from $\U(1) \times \SU(2) \times \SU(3)$ to $\U(2) \times \U(3)$  obtained as the composition
$$
\U(1) \times \SU(2) \times \SU(3) \longrightarrow \U(1) \times \SU(2) \times \U(1) \times \SU(3) \longrightarrow \U(2) \times \U(3) \longrightarrow \U(2) \times \U(3),
$$
where the first map takes $(e^{i\theta},A,B)$ to $(e^{i\theta},A,e^{i \theta}, B)$, the second map takes $(e^{i\alpha},A,e^{i \beta}, B)$ to $(e^{i\alpha}A,e^{i \beta}B)$, and the third map takes $(u,v)$ to $((\det u)u, (\det v)^{-1} v)$. Thus, $\Phi \colon \U(1) \times \SU(2) \times \SU(3) \to \U(2) \times \U(3)$ is simply the map
$$
(e^{i\theta},A,B) \longmapsto (e^{i 3 \theta} A, e^{- i 2 \theta}B).
$$
Since $(\det e^{i 3 \theta} A)(\det e^{- i 2 \theta}B)=e^{i 6 \theta} e^{-i 6 \theta} = 1$, the image of this map is contained in $\uS(\U(2) \times \U(3))$. In fact, it is not hard to check it is equal to $\uS(\U(2) \times \U(3))$. Moreover, the kernel of $\Phi$ corresponds to the subgroup
$$
\ZZ_6 \cong \big\{ \big(e^{2\pi i \frac{k}{6}}, e^{2\pi i \frac{k}{2}} I_2 , e^{2\pi i \frac{k}{3}} I_3 \big) \, \big\vert \, k \in \{1,2,3,4,5,6\}   \big\} \subset \U(1) \times \SU(2) \times \SU(3).
$$
Hence, the homomorphism $\Phi \colon \U(1) \times \SU(2) \times \SU(3) \to \U(2) \times \U(3)$ induces the isomorphic map from $(\U(1) \times \SU(2) \times \SU(3))/\ZZ_6$ onto $\uS(\U(2) \times \U(3))$. As an aside, we remark that the hypercharge group $\U(1)_{\mathnormal{Y}}$ is the image under $\Phi$ of the subgroup
$$
\big\{ \big( e^{i \theta}, I_2, I_3 \big) \, \big\vert \, \theta \in [0,2\pi) \big\} \subset \U(1) \times \SU(2) \times \SU(3),
$$
and that the electromagnetic group $\U(1)_{\mathnormal{Q}}$ is the image under $\Phi$ of the subgroup
$$
\left\{ \left( e^{i \theta},  \left(\begin{array}{cc} e^{ i 3 \theta} & 0 \\ 0 & e^{-i 3 \theta}   \end{array} \right), I_3 \right) \, \Bigg\vert \, \theta \in [0,2\pi) \right\} \subset \U(1) \times \SU(2) \times \SU(3).
$$

It is quite important to note that, in order to describe the full structure of the multiplet representations of the gauge group $\U(2) \times \U(3)$, it is necessary to select a complementary $\U(1)$ to $\uS(\U(2) \times \U(3))$ with respect to a bi-invariant metric on $\U(2) \times \U(3)$. We choose the bi-invariant metric
$$
\langle (u_1,v_2), (u_2,v_2) \rangle = \Re \left[\tr(u_1 u_2^*)\right] + \Re \left[\tr(v_1 v_2^*)\right],
$$
where $(u_1,v_2)$ and $(u_2,v_2)$ are two arbitrary elements of $\U(2) \times \U(3)$ and the superscript `$\ast$' denotes the conjugate transpose. With this choice, the complementary $\U(1)$ to $\uS(\U(2) \times \U(3))$, which we shall write as $\U(1)_{\mathnormal{Z}}$, is easily seen to be
$$
\U(1)_{\mathnormal{Z}} = \big\{ \big( e^{i \theta} I_2, e^{i \theta} I_3 \big) \, \big\vert \, \theta \in [0,2\pi) \big\}. 
$$

\subsection{The interacting-particle representations}\label{Sect:2.3}
In order to find an appropriate representation for the fundamental fermions, we demand that all multiplet representations use only the representation spaces $\CC^1$ and $\CC^3$ for the hyperstrong group $\U(3)$. Of course, $\CC^1$ corresponds to leptons, and $\CC^3$ to quarks. Furthermore, we demand that all multiplet representations use the representation spaces $\CC^1$ and $\CC^2$  for the hyperweak group $\U(2)$.

Let us examine the relevant representations which satisfy the above constraints. Let $\eta$ denote the standard representation of $\U(2)$ on $\CC^2$, and consider the representation $\kappa$ of $\U(2)$ on $\CC^2$ which is obtained by composing the automorphism of $\U(2)$ that sends $u$ to $(\det u) u$ with $\eta$. The former induces the representation $\bigwedge^2 \eta$ of $\U(2)$ on the two-fold exterior power $\bigwedge^2 \CC^2$, on which $u$ acts by multiplication with $\det u$. We also consider, in keeping with the notation introduced in Section \ref{Sect:1.1}, the representation $\lambda_{\mathnormal{L}}^3$ of the $\U(1)_{\mathnormal{L}}$ subgroup of $\U(2)$ on $\CC^1_{3}$. Thus, the representation $\kappa_{\mathnormal{L}}$ of $\U(1)_{\mathnormal{L}} \times \SU(2)$ on $\CC^2$, which is obtained by composing the covering homomorphism $r_{\mathnormal{L}} \colon \U(1)_{\mathnormal{L}} \times \SU(2) \to \U(2)$ with the representation $\kappa$ of $\U(2)$ on $\CC^2$, is equivalent to the representation $\lambda_{\mathnormal{L}}^3 \otimes \tau$ of $\U(1)_{\mathnormal{L}} \times \SU(2)$ on $\CC^1_3 \otimes \CC^2$. To see this, let $T \colon \CC^1_3 \otimes \CC^2 \to \CC^2$ denote the linear isomorphism defined by $T(z \otimes w)=z w$. Then, for each $(e^{i\alpha},A) \in \U(1)_{\mathnormal{L}} \times \SU(2)$ and for an arbitrary element $z \otimes w \in \CC^1_3 \otimes \CC^2$, 
$$
[\kappa_{\mathnormal{L}}(e^{i\alpha},A) \circ T](z \otimes w)=[\kappa(e^{i\alpha}A)](zw)=e^{i 3 \alpha} A z w=e^{i 3 \alpha}z A w,
$$
and 
$$
[T \circ (\lambda^3_{\mathnormal{L}} \otimes \tau)(e^{i\alpha},A)](z \otimes w)=T(e^{i 3 \alpha} z \otimes A w)=e^{i 3 \alpha}z A w.
$$
This means that for each $h \in \U(1)_{\mathnormal{L}} \times \SU(2)$, we obtain
$$
\kappa_{\mathnormal{L}}(h)\circ T = T \circ (\lambda^3_{\mathnormal{L}} \otimes \tau)(h),
$$
which implies the assertion.

In an analogous manner, let $\zeta$ denote the standard representation of $\U(3)$ on $\CC^3$, and consider the representation $\iota$ of $\U(3)$ on $\CC^3$ which is obtained by composing the automorphism of $\U(3)$ that sends $v$ to $(\det v)^{-1} v$ with $\zeta$. The former induces the representation $\bigwedge^3 \zeta$ on the three-fold exterior power $\bigwedge^3 \CC^3$, on which $v$ acts by multiplication with $\det v$. We may consider, too, the representation $\lambda^{-2}_{\mathnormal{C}}$ of the $\U(1)_{\mathnormal{C}}$ subgroup of $\U(2)$ on $\CC^1_{-2}$. Then, just as above, we find that the representation $\iota_{\mathnormal{C}}$ of $\U(1)_{\mathnormal{C}} \times \SU(3)$ on $\CC^3$, which is obtained by composing the covering homomorphism $r_{\mathnormal{C}} \colon \U(1)_{\mathnormal{C}} \times \SU(3) \to \U(3)$ with the representation $\iota$ of $\U(3)$ on $\CC^3$, is equivalent to the representation $\lambda_{\mathnormal{C}}^{-2}\otimes \rho$ of $\U(1)_{\mathnormal{C}} \times \SU(3)$ on $\CC^1_{-2}\otimes \CC^3$.

In view of Section~\ref{Sect:2.1}, the simplest prescription is to demand that all fundamental fermions in our model are expressed as multiplets by the representations of $\U(2) \times \U(3)$ outlined in Table~\ref{tab:4}.
\begin{table}[ht]
\begin{tabular}{lcc}
\hline
\hline
Name                     & Symbol                 & Representation                              
	 \\
	 \hline 
	 \\
         Left-handed leptons      & $\left(\begin{array}{c} \nu_{\mathnormal{L}} \\ e_{\mathnormal{L}} \end{array}\right)$                 & $\bigwedge^1\eta\otimes \bigwedge^3\zeta$ \\
         \\                                                               
         Left-handed quarks       & $\left(\begin{array}{c} u_{\mathnormal{L}} \\ d_{\mathnormal{L}} \end{array}\right)$      & $\bigwedge^1\eta\otimes \bigwedge^1\zeta$ \\
         \\                                                               
         Right-handed neutrino    & $\nu_{\mathnormal{R}}$                & $\bigwedge^2\eta\otimes \bigwedge^3\zeta$ \\
	 \\                                                               
         Right-handed electron    & $e_{\mathnormal{R}}$                & $\bigwedge^0\eta\otimes \bigwedge^3\zeta$ \\
	 \\                                                               
         Right-handed up quark   & $u_{\mathnormal{R}}$  &  $\bigwedge^2\eta\otimes \bigwedge^1\zeta$ \\
	 \\                                                               
         Right-handed down quark & $d_{\mathnormal{R}}$  & $\bigwedge^0\eta\otimes \bigwedge^1\zeta$ \\ \\
	 \hline
	\end{tabular}
	\vspace{-5pt}
\caption{Fundamental fermions as representations of
$\U(2) \times \U(3)$} \label{tab:4}
\end{table}

As with the standard model representations, parity violation is incorporated by assigning the left-  and right-handed components of the fermions to different group representations. Again, there is a tacit understanding that each of the representations in Table~\ref{tab:4} is tensored with the left- or right- handed spinor representation of $\SL(2,\CC)$, as appropriate.  

On the other hand, as the discussion above makes clear, every representation in Table~\ref{tab:4} can be lifted to a representation of $\U(1)_{\mathnormal{L}} \times \SU(2) \times \U(1)_{\mathnormal{C}} \times \SU(3)$. These representations are given in Table~\ref{tab:5} below. 
\begin{table}[b]
\begin{tabular}{lcc}
\hline
\hline
Name                     & Symbol                 & Representation                              
	 \\
	 \hline 
	 \\
         Left-handed leptons      & $\left(\begin{array}{c} \nu_{\mathnormal{L}} \\ e_{\mathnormal{L}} \end{array}\right)$                 & $\bigwedge^1(\lambda_{\mathnormal{L}}^{3}\otimes \tau)\otimes \bigwedge^3(\lambda_{\mathnormal{C}}^{-2}\otimes \rho)$ \\
         \\                                                               
         Left-handed quarks       & $\left(\begin{array}{c} u_{\mathnormal{L}} \\ d_{\mathnormal{L}} \end{array}\right)$      & $\bigwedge^1(\lambda_{\mathnormal{L}}^{3}\otimes \tau)\otimes \bigwedge^1(\lambda_{\mathnormal{C}}^{-2}\otimes \rho)$ \\
         \\                                                               
         Right-handed neutrino    & $\nu_{\mathnormal{R}}$                & $\bigwedge^2(\lambda_{\mathnormal{L}}^{3}\otimes \tau)\otimes \bigwedge^3(\lambda_{\mathnormal{C}}^{-2}\otimes \rho)$ \\
	 \\                                                               
         Right-handed electron    & $e_{\mathnormal{R}}$                & $\bigwedge^0(\lambda_{\mathnormal{L}}^{3}\otimes \tau)\otimes \bigwedge^3(\lambda_{\mathnormal{C}}^{-2}\otimes \rho)$ \\
	 \\                                                               
         Right-handed up quark   & $u_{\mathnormal{R}}$  &  $\bigwedge^2(\lambda_{\mathnormal{L}}^{3}\otimes \tau)\otimes \bigwedge^1(\lambda_{\mathnormal{C}}^{-2}\otimes \rho)$ \\
	 \\                                                               
         Right-handed down quark & $d_{\mathnormal{R}}$  & $\bigwedge^0(\lambda_{\mathnormal{L}}^{3}\otimes \tau)\otimes \bigwedge^1(\lambda_{\mathnormal{C}}^{-2}\otimes \rho)$ \\ \\
	 \hline
	\end{tabular}
	\vspace{-5pt}
\caption{Fundamental fermions as representations of
$\U(1)_{\mathnormal{L}}  \times \SU(2) \times \U(1)_{\mathnormal{C}} \times \SU(3)$} \label{tab:5}
\end{table}

If we further restrict the representations of Table~\ref{tab:5} to the image of $\U(1) \times \SU(2) \times \SU(3)$ under the inclusion $\U(1) \times \SU(2) \times \SU(3) \to \U(1) \times \SU(2) \times \U(1) \times \SU(3)$ that takes $(e^{i\theta},A,B)$ to $(e^{i\theta},A,e^{i\theta},B)$, we get exactly the fermionic multiplet representations in the standard model as displayed in Table~\ref{tab:3}. Thus, we see that Table~\ref{tab:4} is compatible with the usual classification of the leptons and quarks into left- and right-handed multiplets.

Now we shall address two key points: first, the apparent preference for the representations $\eta$ and $\zeta$ of $\U(2)$ and $\U(3)$, and for the odd part of the exterior algebra of $\zeta$, and second, the question of whether or not our model should have additional fermions beyond those contained in the standard model. 

For the first point, notice that the most ``democratic'' representation of $\U(2) \times \U(3)$ within the given range of tensorial orders, that takes into account all the possible chirality assignments, is
$$
\left( \textstyle \bigwedge^{\bullet} \eta \oplus \textstyle \bigwedge^{\bullet} \overline{\eta} \right) \otimes \left( \textstyle \bigwedge^{\bullet} \zeta \oplus \textstyle \bigwedge^{\bullet} \overline{\zeta} \right) \otimes \sigma.
$$
Of course, this representation contains the interacting particle representation for the fundamental fermions of Table~\ref{tab:4},
\begin{align*}
&\left[\textstyle \bigwedge^{\mathrm{odd}} \eta \otimes \textstyle \bigwedge^{\mathrm{odd}} \zeta \otimes \sigma_L \right] \oplus \left[\textstyle \bigwedge^{\mathrm{even}} \eta \otimes \textstyle \bigwedge^{\mathrm{odd}} \zeta \otimes \sigma_R \right] \\
&\quad =\left(\textstyle\bigwedge^1 \eta \otimes \bigwedge^3 \zeta  \otimes \sigma_L\right)  \oplus \left(\textstyle\bigwedge^1 \eta \otimes \bigwedge^1 \zeta \otimes \sigma_L \right) \oplus \left(\textstyle\bigwedge^2 \eta \otimes \bigwedge^3 \zeta \otimes \sigma_R \right) \\& \quad\quad \oplus \left(\textstyle\bigwedge^0 \eta \otimes \bigwedge^3 \zeta  \otimes \sigma_R\right) \oplus \left(\textstyle\bigwedge^2 \eta \otimes \bigwedge^1 \zeta \otimes \sigma_R \right) \oplus \left(\textstyle\bigwedge^0 \eta \otimes \bigwedge^1 \zeta \otimes \sigma_R\right),
\end{align*}
as well as fifteen additional blocks. One of these blocks, namely 
$$
\left[\textstyle \bigwedge^{\mathrm{odd}} \overline{\eta} \otimes \textstyle \bigwedge^{\mathrm{odd}} \overline{\zeta} \otimes \sigma_R \right] \oplus \left[\textstyle \bigwedge^{\mathrm{even}} \overline{\eta} \otimes \textstyle \bigwedge^{\mathrm{odd}} \overline{\zeta}\otimes \sigma_L \right],
$$
plays the role of the interacting antiparticle representation.

For the second point, recall that realistic extensions of the standard model require additional constraints to be fulfilled. Perhaps the most stringent of them all is that of anomaly cancellation. As will be shown in Section~\ref{Sect:2.5} below, such a constraint dictates the presence of additional fermions. This is simply a consequence of the fact that the fundamental fermions listed in Table~\ref{tab:4} may be charged under the complementary $\U(1)_{\mathnormal{Z}} \subset \U(2) \times \U(3)$ introduced towards the end of the preceding section. A collection of new fermions which, together with the fundamental fermions in Table~\ref{tab:4}, form an anomaly free set of fermions are recorded in Table~\ref{tab:5a}, where $\overline{\eta}$ and $\overline{\zeta}$ correspond to the conjugate representations of $\eta$ and $\zeta$, respectively.
\begin{table}[ht]
\begin{tabular}{lcc}
\hline
\hline
 Name & Symbol                 & Representation                              
	 \\
	 \hline 
	 \\
 Exotic left-handed leptons &        $\left(\begin{array}{c} E_{\mathnormal{L}} \\ N_{\mathnormal{L}} \end{array}\right)$                 & $\bigwedge^1\overline{\eta}\otimes \bigwedge^0\overline{\zeta}$ \\
         \\                                                               
 Exotic left-handed quarks  &      $\left(\begin{array}{c} D_{\mathnormal{L}} \\ U_{\mathnormal{L}} \end{array}\right)$      & $\bigwedge^1\overline{\eta}\otimes \bigwedge^2\overline{\zeta}$ \\
         \\                                                               
Exotic right-handed neutrino &      $N_{\mathnormal{R}}$                & $\bigwedge^0\overline{\eta}\otimes \bigwedge^0\overline{\zeta}$ \\
	 \\                                                               
   Exotic right-handed electron &       $E_{\mathnormal{R}}$                & $\bigwedge^2\overline{\eta}\otimes \bigwedge^0\overline{\zeta}$ \\
	 \\                                                               
   Exotic right-handed up quark &       $U_{\mathnormal{R}} $  &  $\bigwedge^0\overline{\eta}\otimes \bigwedge^2\overline{\zeta}$ \\
	 \\                                                               
  Exotic right-handed down quark &    $D_{\mathnormal{R}}$  & $\bigwedge^2\overline{\eta}\otimes \bigwedge^2\overline{\zeta}$ \\ \\
	 \hline
	\end{tabular}
	\vspace{-5pt}
\caption{Exotic fermions as representations of
$\U(2) \times \U(3)$} \label{tab:5a}
\end{table}
Thus their interacting particle representation corresponds to the block
\begin{align*}
&\left[\textstyle \bigwedge^{\mathrm{odd}} \overline{\eta} \otimes \textstyle \bigwedge^{\mathrm{even}} \overline{\zeta}  \otimes \sigma_L\right] \oplus \left[\textstyle \bigwedge^{\mathrm{even}} \overline{\eta} \otimes \textstyle \bigwedge^{\mathrm{even}} \overline{\zeta}  \otimes \sigma_R\right] \\
 &\quad=\left(\textstyle\bigwedge^1 \overline{\eta} \otimes \bigwedge^0 \overline{\zeta} \otimes \sigma_L \right)  \oplus \left(\textstyle\bigwedge^1 \overline{\eta} \otimes \bigwedge^2 \overline{\zeta} \otimes \sigma_L \right) \oplus \left(\textstyle\bigwedge^0 \overline{\eta} \otimes \bigwedge^0 \overline{\zeta} \otimes \sigma_R \right) \\& \quad \quad\oplus \left(\textstyle\bigwedge^2 \overline{\eta} \otimes \bigwedge^0 \overline{\zeta} \otimes \sigma_R \right) \oplus \left(\textstyle\bigwedge^0 \overline{\eta} \otimes \bigwedge^2 \overline{\zeta} \otimes \sigma_R \right) \oplus \left(\textstyle\bigwedge^2 \overline{\eta} \otimes \bigwedge^2 \overline{\zeta} \otimes \sigma_R \right),
\end{align*}
and the associated interacting antiparticle representation to the block 
$$
\left[\textstyle \bigwedge^{\mathrm{odd}} \eta \otimes \textstyle \bigwedge^{\mathrm{even}} \zeta  \otimes \sigma_R\right] \oplus \left[\textstyle \bigwedge^{\mathrm{even}} \eta \otimes \textstyle \bigwedge^{\mathrm{even}} \zeta  \otimes \sigma_L\right] .
$$
Other choices of extra fermions are also possible. Nevertheless, here, we have chosen the unique block containing the ``exotics'' that best emulate the fundamental fermions of Table~\ref{tab:4} together with their chiralities. For all these other choices, the calculations discussed below and in the next section can be carried out in a similar fashion. 

Using an argument entirely analogous to the one leading to Table~\ref{tab:5}, every representation in Table~\ref{tab:5a} can be lifted to a representation of $\U(1)_{\mathnormal{L}} \times \SU(2) \times \U(1)_{\mathnormal{C}} \times \SU(3)$. These representations are shown in Table~\ref{tab:5b}, where we have used the fact that the standard representation $\tau$ of $\SU(2)$ on $\CC^2$ is self-dual.
\begin{table}[ht]
\begin{tabular}{lcc}
\hline
\hline
Name & Symbol                 & Representation                              
	 \\
	 \hline 
	 \\
   Exotic left-handed leptons   &   $\left(\begin{array}{c} E_{\mathnormal{L}} \\ N_{\mathnormal{L}} \end{array}\right)$                 & $\bigwedge^1(\lambda^{-3}_{\mathnormal{L}} \otimes \tau)\otimes \bigwedge^0(\lambda^{2}_{\mathnormal{C}} \otimes \overline{\rho})$ \\
         \\                                                               
 Exotic left-handed quarks    &    $\left(\begin{array}{c} D_{\mathnormal{L}} \\ U_{\mathnormal{L}} \end{array}\right)$      & $\bigwedge^1(\lambda^{-3}_{\mathnormal{L}} \otimes \tau)\otimes \bigwedge^2(\lambda^{2}_{\mathnormal{C}} \otimes \overline{\rho})$ \\
         \\                                                               
 Exotic right-handed neutrino   &     $N_{\mathnormal{R}}$                & $\bigwedge^0(\lambda^{-3}_{\mathnormal{L}} \otimes \tau)\otimes \bigwedge^0(\lambda^{2}_{\mathnormal{C}} \otimes \overline{\rho})$ \\
	 \\                                                               
  Exotic right-handed electron &    $E_{\mathnormal{R}}$                & $\bigwedge^2(\lambda^{-3}_{\mathnormal{L}} \otimes \tau)\otimes \bigwedge^0(\lambda^{2}_{\mathnormal{C}} \otimes \overline{\rho})$ \\
	 \\                                                               
 Exotic right-handed up quark &   $U_{\mathnormal{R}} $  &  $\bigwedge^0(\lambda^{-3}_{\mathnormal{L}} \otimes \tau)\otimes \bigwedge^2(\lambda^{2}_{\mathnormal{C}} \otimes \overline{\rho})$ \\
	 \\                                                               
    Exotic right-handed down quark &       $D_{\mathnormal{R}}$  & $\bigwedge^2(\lambda^{-3}_{\mathnormal{L}} \otimes \tau)\otimes \bigwedge^2(\lambda^{2}_{\mathnormal{C}} \otimes \overline{\rho})$ \\ \\
	 \hline
	\end{tabular}
	\vspace{-5pt}
\caption{Exotic fermions as representations of
$\U(1)_{\mathnormal{L}} \times \SU(2) \times \U(1)_{\mathnormal{C}} \times \SU(3)$} \label{tab:5b}
\end{table}

To conclude this section, we briefly describe the hypercharge values of the different fermionic multiplets. We shall first consider the hypercharge values with respect to the hypercharge group $\U(1)_{\mathnormal{Y}}$. By definition, the hypercharge operator is an appropriate normalisation of the generator of $\U(1)_{\mathnormal{Y}}$. Let $Y^{\mathnormal{L}}$ (respectively, $Y^{\mathnormal{R}}$) be such a generator regarded as a matrix acting on the left-handed multiplets (respectively, right-handed multiplets). Looking closely at Tables~\ref{tab:4} and~\ref{tab:5a}, one easily finds
\begin{gather*}
Y^{\mathnormal{L}} = \left( \scalemath{1}{ \begin{array}{cccccccc} -3  & 0 & 0 & 0 & 0 & 0 & 0 & 0 \\
0 & -3 & 0 & 0 & 0 & 0 & 0 & 0\\ 
0 & 0 & 1 & 0 & 0 & 0 & 0 & 0\\
0 & 0 & 0 & 1 & 0 & 0 & 0 & 0  \\
0 & 0 & 0 & 0 & 1 & 0 & 0 & 0\\
0 & 0 & 0 & 0 & 0 & 1 & 0  & 0\\
0 & 0 & 0 & 0 & 0 & 0 & 1 & 0 \\
0 & 0 & 0 & 0 & 0 & 0 & 0 & 1 \end{array} }\right) \oplus \left( \scalemath{1}{\begin{array}{cccccccc} -3  & 0 & 0 & 0 & 0 & 0 & 0 & 0 \\
0 & -3 & 0 & 0 & 0 & 0 & 0 & 0\\ 
0 & 0 & 1 & 0 & 0 & 0 & 0 & 0\\
0 & 0 & 0 & 1 & 0 & 0 & 0 & 0  \\
0 & 0 & 0 & 0 & 1 & 0 & 0 & 0\\
0 & 0 & 0 & 0 & 0 & 1 & 0  & 0\\
0 & 0 & 0 & 0 & 0 & 0 & 1 & 0 \\
0 & 0 & 0 & 0 & 0 & 0 & 0 & 1 \end{array}}\right)
\end{gather*}
and 
$$
Y^{\mathnormal{R}} = \left( \scalemath{1}{\begin{array}{cccccccc} 0  & 0 & 0 & 0 & 0 & 0 & 0 & 0 \\
0 & -6 & 0 & 0 & 0 & 0 & 0 & 0\\ 
0 & 0 & 4 & 0 & 0 & 0 & 0 & 0\\
0 & 0 & 0 & 4 & 0 & 0 & 0 & 0  \\
0 & 0 & 0 & 0 & 4 & 0 & 0 & 0\\
0 & 0 & 0 & 0 & 0 & -2 & 0  & 0\\
0 & 0 & 0 & 0 & 0 & 0 & -2 & 0 \\
0 & 0 & 0 & 0 & 0 & 0 & 0 & -2 \end{array}}\right) \oplus \left( \scalemath{1}{\begin{array}{cccccccc} 0  & 0 & 0 & 0 & 0 & 0 & 0 & 0 \\
0 & -6 & 0 & 0 & 0 & 0 & 0 & 0\\ 
0 & 0 & 4 & 0 & 0 & 0 & 0 & 0\\
0 & 0 & 0 & 4 & 0 & 0 & 0 & 0  \\
0 & 0 & 0 & 0 & 4 & 0 & 0 & 0\\
0 & 0 & 0 & 0 & 0 & -2 & 0  & 0\\
0 & 0 & 0 & 0 & 0 & 0 & -2 & 0 \\
0 & 0 & 0 & 0 & 0 & 0 & 0 & -2 \end{array}}\right),
$$
where, in both cases, the first block summand refers to the fundamental fermions, and the second to the exotic fermions. Thus we have a simple rule for the hypercharge $Y(\psi)$ of a fermionic multiplet $\psi$ for the group $\U(1)_{\mathnormal{Y}}$, namely
$$
Y(\psi) =y\left[3 \ord_2(\psi) - 2 \ord_3(\psi)\right],
$$
where $y$ is a real nonzero normalisation constant and where $\ord_2(\psi)$ and $\ord_3(\psi)$ are the tensorial orders of the $\U(2)$ and $\U(3)$ representations, respectively (cf.~\cite{Hucks91}). In order to agree with the observed hypercharges of the quarks and leptons, we take $y=\frac{1}{3}$. Next, we shall consider the hypercharge values with respect to the complementary $\U(1)_{\mathnormal{Z}}$ to $\uS(\U(2) \times \U(3))$. The hypercharge operator is, by definition, an appropriate normalisation of the generator of $\U(1)_{\mathnormal{Z}}$. So let $Z^{\mathnormal{L}}$ (respectively, $Z^{\mathnormal{R}}$) be this generator regarded as a matrix acting on the left-handed multiplets (respectively, right-handed multiplets). Using the representations from Tables~\ref{tab:4} and~\ref{tab:5a}, one obtains 
$$
Z^{\mathnormal{L}} = \left( \scalemath{1}{\begin{array}{cccccccc} 4  & 0 & 0 & 0 & 0 & 0 & 0 & 0 \\
0 & 4 & 0 & 0 & 0 & 0 & 0 & 0\\ 
0 & 0 & 2 & 0 & 0 & 0 & 0 & 0\\
0 & 0 & 0 & 2 & 0 & 0 & 0 & 0  \\
0 & 0 & 0 & 0 & 2 & 0 & 0 & 0\\
0 & 0 & 0 & 0 & 0 & 2 & 0  & 0\\
0 & 0 & 0 & 0 & 0 & 0 & 2 & 0 \\
0 & 0 & 0 & 0 & 0 & 0 & 0 & 2 \end{array}}\right) \oplus \left( \scalemath{1}{\begin{array}{cccccccc} -1  & 0 & 0 & 0 & 0 & 0 & 0 & 0 \\
0 & -1 & 0 & 0 & 0 & 0 & 0 & 0\\ 
0 & 0 & -3 & 0 & 0 & 0 & 0 & 0\\
0 & 0 & 0 & -3 & 0 & 0 & 0 & 0  \\
0 & 0 & 0 & 0 & -3 & 0 & 0 & 0\\
0 & 0 & 0 & 0 & 0 & -3 & 0  & 0\\
0 & 0 & 0 & 0 & 0 & 0 & -3 & 0 \\
0 & 0 & 0 & 0 & 0 & 0 & 0 & -3 \end{array}}\right)
$$
and 
$$
Z^{\mathnormal{R}} = \left( \scalemath{1}{\begin{array}{cccccccc} 5  & 0 & 0 & 0 & 0 & 0 & 0 & 0 \\
0 & 3 & 0 & 0 & 0 & 0 & 0 & 0\\ 
0 & 0 & 3 & 0 & 0 & 0 & 0 & 0\\
0 & 0 & 0 & 3 & 0 & 0 & 0 & 0  \\
0 & 0 & 0 & 0 & 3 & 0 & 0 & 0\\
0 & 0 & 0 & 0 & 0 & 1 & 0  & 0\\
0 & 0 & 0 & 0 & 0 & 0 & 1 & 0 \\
0 & 0 & 0 & 0 & 0 & 0 & 0 & 1 \end{array}}\right) \oplus \left( \scalemath{1}{\begin{array}{cccccccc} 0  & 0 & 0 & 0 & 0 & 0 & 0 & 0 \\
0 & -2 & 0 & 0 & 0 & 0 & 0 & 0\\ 
0 & 0 & -2 & 0 & 0 & 0 & 0 & 0\\
0 & 0 & 0 & -2 & 0 & 0 & 0 & 0  \\
0 & 0 & 0 & 0 & -2 & 0 & 0 & 0\\
0 & 0 & 0 & 0 & 0 & -4 & 0  & 0\\
0 & 0 & 0 & 0 & 0 & 0 & -4 & 0 \\
0 & 0 & 0 & 0 & 0 & 0 & 0 & -4 \end{array}}\right).
$$
The hypercharge $Z(\psi)$ of a fermionic multiplet for the group $\U(1)_{\mathnormal{Z}}$ then becomes
$$
Z(\psi)= z \left[ \ord_2(\psi) + \ord_3(\psi) \right],
$$   
where $z$ is a normalisation constant. We shall take $z = \frac{1}{3}$. The resulting hypercharges for both the fundamental and exotic fermions are presented in Table~\ref{tab:6}.  
\begin{table}[ht]
\begin{tabular}{ccc | ccc}
\hline
\hline
 Symbol                 & Hypercharge  $Y$   & Hypercharge $Z$      & Symbol                 & Hypercharge $Y$   & Hypercharge $Z$                            
	 \\
	 \hline 
& & & & &	 \\
         $\left(\begin{array}{c} \nu_{\mathnormal{L}} \\ e_{\mathnormal{L}} \end{array}\right)$                 & $-1$ &  $\phantom{-}\frac{4}{3}$ &   $\left(\begin{array}{c} E_{\mathnormal{L}} \\ N_{\mathnormal{L}} \end{array}\right)$                 & $-1$ &  $-\frac{1}{3}$\\
  & & & & &       \\                                                               
         $\left(\begin{array}{c} u_{\mathnormal{L}} \\ d_{\mathnormal{L}} \end{array}\right)$      & $\phantom{-}\frac{1}{3}$ & $\phantom{-}\frac{2}{3}$ &   $\left(\begin{array}{c} D_{\mathnormal{L}} \\ U_{\mathnormal{L}} \end{array}\right)$      & $\phantom{-}\frac{1}{3}$ & $-1$ \\
   & & & & &      \\                                                               
          $\nu_{\mathnormal{R}}$                & $\phantom{-}0$  & $\phantom{-}\frac{5}{3}$ &   $N_{\mathnormal{R}}$                & $\phantom{-}0$  & $\phantom{-}0$\\
& & & & &	 \\                                                               
          $e_{\mathnormal{R}}$                & $-2$ & $\phantom{-}1$ &    $E_{\mathnormal{R}}$                & $-2$ & $-\frac{2}{3}$  \\
& & & & &	 \\                                                               
          $u_{\mathnormal{R}}$  &  $\phantom{-}\frac{4}{3}$ & $\phantom{-}1$ &   $U_{\mathnormal{R}}$  &  $\phantom{-}\frac{4}{3}$ & $-\frac{2}{3}$ \\
& & & & &	 \\                                                               
          $d_{\mathnormal{R}}$  & $-\frac{2}{3}$ & $\phantom{-}\frac{1}{3}$ &   $D_{\mathnormal{R}}$  & $-\frac{2}{3}$ & $-\frac{4}{3}$ \\ 
  & & & & &       \\                                                    
	 \hline
	\end{tabular}
	\vspace{-5pt}
\caption{Hypercharge assignment for the fundamental and exotic fermions} \label{tab:6}
\end{table}

\subsection{Spontaneous symmetry breaking of $\U(2) \times \U(3)$ to $\U(3)_{\mathnormal{Q}}$}\label{Sect:2.4}
As we have already indicated, under spontaneous symmetry breaking, the representations for the fermionic multiplets of Tables~\ref{tab:4} and \ref{tab:5a} decompose into direct sums of electrostrong $\U(3)_{\mathnormal{Q}}$ representations. These electrostrong representations correspond to the ``physical'' particles. Let us proceed to describe them. 

We first deal with the fundamental fermions of Table~\ref{tab:4}. Consider the representation $\xi^{\pm}$ of $\U(3)_{\mathnormal{Q}}$ on $\CC^3$ which sends $v$ to $(\det v)^{\pm 1} v$ acting on $\CC^3$. This representation induces the representations $\bigwedge^3 \xi^{\pm}$ of $\U(3)_{\mathnormal{Q}}$ on the three-fold exterior power $\bigwedge^3 \CC^3$, in which $v$ acts as $(\det v)^{\pm 3 + 1}$ upon $\bigwedge^3 \CC^3$. We also consider the trivial representation $1_{\U(3)_{\mathnormal{Q}}}$ of $\U(3)_{\mathnormal{Q}}$. After the symmetry breaking has taken place, all the elementary fermions are assigned to the representations listed in Table~\ref{tab:7}. 
\begin{table}[ht]
\begin{tabular}{lcc}
\hline
\hline
Name                     & Symbol                 & Representation                              
	 \\
	 \hline 
	 \\
         Neutrino      & $\nu_e$                 & $1_{\U(3)_{\mathnormal{Q}}}$ \\
         \\                                                               
         Electron    & $e$                & $\bigwedge^3\xi^{-}$ \\
	 \\                                                               
            Up quark   & $u$  &  $\xi^{+}$ \\
	 \\                                                               
         Down quark & $d$  & $\xi^{-}$ \\ \\
	 \hline
	\end{tabular}
	\vspace{-5pt}
\caption{Elementary fermions as electrostrong $\U(3)_{\mathnormal{Q}}$ representations} \label{tab:7}
\end{table}

Each of the representations in Table~\ref{tab:7} must be tensored with the Dirac spinor representation $\sigma=\sigma_{\mathnormal{L}} \oplus \sigma_{\mathnormal{R}}$ of $\SL(2,\CC)$. In order to justify this we discuss how, under symmetry breaking, the representations for the fundamental fermions of Table~\ref{tab:4} break into direct sums of representations housing the elementary fermions in Table~\ref{tab:7}.

The restriction of the representation $\bigwedge^1 \eta \otimes \bigwedge^3 \zeta$ to $\U(3)_{\mathnormal{Q}}$ decomposes as a direct sum of two irreducible representations of $\U(3)_{\mathnormal{Q}}$. 
To be precise,
\begin{align*}
&\left(\textstyle\bigwedge^1 \eta \otimes \bigwedge^3 \zeta\right)\left(\left(\begin{array}{cc} (\det v)^{2} & 0 \\ 0 & 1   \end{array} \right), (\det v)^{-1}v \right) \\
&\quad=\left(\begin{array}{cc} (\det v)^{2} & 0 \\ 0 & 1   \end{array} \right) \otimes (\det v)^{-2} =\left(\begin{array}{cc} 1 & 0 \\ 0 & (\det v)^{-2}   \end{array} \right) \otimes 1=(1 \otimes 1) \oplus ((\det v)^{-2} \otimes 1).
\end{align*}
These two representations are equivalent to the representations $1_{\U(3)_{\mathnormal{Q}}}$ and $\bigwedge^3 \xi^{-}$, respectively. Hence, under spontaneous symmetry breaking, the representation $\bigwedge^1 \eta \otimes \bigwedge^3 \zeta$ reduces to $1_{\U(3)_{\mathnormal{Q}}} \oplus \bigwedge^3 \xi^{-}$. Thus, $\bigwedge^1 \eta \otimes \bigwedge^3 \zeta \otimes \sigma_{\mathnormal{L}}$ reduces to $(1_{\U(3)_{\mathnormal{Q}}} \oplus \bigwedge^3 \xi^{-}) \otimes \sigma_{\mathnormal{L}}$.

In a similar way, the restriction of the representation $\bigwedge^1 \eta \otimes \bigwedge^1 \zeta$ to $\U(3)_{\mathnormal{Q}}$ decomposes as a direct sum of two irreducible representations of $\U(3)_{\mathnormal{Q}}$. Indeed,
\begin{align*}
&\left(\textstyle\bigwedge^1 \eta \otimes \bigwedge^1 \zeta\right)\left(\left(\begin{array}{cc} (\det v)^{2} & 0 \\ 0 & 1   \end{array} \right), (\det v)^{-1}v \right) \\
&\quad=\left(\begin{array}{cc} (\det v)^{2} & 0 \\ 0 & 1   \end{array} \right) \otimes (\det v)^{-1}v =\left(\begin{array}{cc} \det v & 0 \\ 0 & (\det v)^{-1} \end{array} \right) \otimes v=(\det v \otimes v) \oplus ((\det v)^{-1} \otimes v).
\end{align*}
These two representations are equivalent to the representations $\xi^{+}$ and $\xi^{-}$, respectively. Hence, $\bigwedge^1 \eta \otimes \bigwedge^1 \zeta \otimes \sigma_{\mathnormal{L}}$ reduces to $(\xi^{+}\oplus \xi^{-}) \otimes \sigma_{\mathnormal{L}}$.

On the other hand, the restriction of the representation $\bigwedge^2 \eta \otimes \bigwedge^3 \zeta$ to $\U(3)_{\mathnormal{Q}}$ is equivalent to the representation $1_{\U(3)_{\mathnormal{Q}}}$, so that $\bigwedge^2 \eta \otimes \bigwedge^3 \zeta \otimes \sigma_{\mathnormal{R}}$ reduces to $\sigma_{\mathnormal{R}}$.

It is also the case that the restriction of the representation $\bigwedge^0 \eta \otimes \bigwedge^3 \zeta$ to $\U(3)_{\mathnormal{Q}}$ is equivalent to the representation $\bigwedge^3 \xi^{-}$, so that $\bigwedge^0 \eta \otimes \bigwedge^3 \zeta \otimes \sigma_{\mathnormal{R}}$ reduces to $\bigwedge^3 \xi^{-} \otimes \sigma_{\mathnormal{R}}$.  

In contrast, the restriction of the representation $\bigwedge^2 \eta \otimes \bigwedge^1 \zeta$ to $\U(3)_{\mathnormal{Q}}$ is equivalent to the representation $\xi^{+}$. Hence, $\bigwedge^2 \eta \otimes \bigwedge^1 \zeta \otimes \sigma_{\mathnormal{R}}$ reduces to $\xi^{+}\otimes \sigma_{\mathnormal{R}}$.

Finally, the restriction of the representation $\bigwedge^0 \eta \otimes \bigwedge^1 \zeta$ to $\U(3)_{\mathnormal{Q}}$ is equivalent to the representation $\xi^{-}$. As a result, $\bigwedge^0 \eta \otimes \bigwedge^1 \zeta \otimes \sigma_{\mathnormal{R}}$ reduces to $\xi^{-}\otimes \sigma_{\mathnormal{R}}$.

Therefore, the $\U(2) \times \U(3)$ model interacting particle representation for the fundamental fermions,  
\begin{align*}
&\left(\textstyle\bigwedge^1 \eta \otimes \bigwedge^3 \zeta \otimes \sigma_{\mathnormal{L}} \right)  \oplus \left(\textstyle\bigwedge^1 \eta \otimes \bigwedge^1 \zeta \otimes \sigma_{\mathnormal{L}} \right) \oplus \left(\textstyle\bigwedge^2 \eta \otimes \bigwedge^3 \zeta \otimes \sigma_{\mathnormal{R}} \right) \\& \quad \oplus \left(\textstyle\bigwedge^0 \eta \otimes \bigwedge^3 \zeta \otimes \sigma_{\mathnormal{R}} \right) \oplus \left(\textstyle\bigwedge^2 \eta \otimes \bigwedge^1 \zeta \otimes \sigma_{\mathnormal{R}} \right) \oplus \left(\textstyle\bigwedge^0 \eta \otimes \bigwedge^1 \zeta \otimes \sigma_{\mathnormal{R}} \right),
\end{align*}
reduces to
\begin{align*}
&\left[ \left(1_{\U(3)_{\mathnormal{Q}}} \oplus \textstyle\bigwedge^3 \xi^{-} \right)  \otimes \sigma_{\mathnormal{L}} \right] \oplus \left[ \left(\xi^{+}\oplus \xi^{-} \right)  \otimes \sigma_{\mathnormal{L}} \right] \oplus \sigma_{\mathnormal{R}}  \oplus \left(\textstyle\bigwedge^3 \xi^{-} \otimes \sigma_{\mathnormal{R}} \right) \oplus \left(\xi^{+}\otimes \sigma_{\mathnormal{R}} \right) \oplus \left(\xi^{-} \otimes \sigma_{\mathnormal{R}} \right) \\
& \quad = \sigma \oplus \left( \textstyle\bigwedge^3 \xi^{-} \otimes \sigma \right) \oplus \left( \xi^{+}\otimes \sigma \right) \oplus \left( \xi^{-}\otimes \sigma \right).
\end{align*}
The latter is precisely the electrostrong $\U(3)_{\mathnormal{Q}}$ interacting-particle representation for the elementary fermions recorded in Table~\ref{tab:7}: $\sigma$ represents the electron-neutrino $\nu_e$, $\bigwedge^3 \xi^{-} \otimes \sigma$ represents the electron, $\xi^{+}\otimes \sigma$ represents the up quark, and $\xi^{-}\otimes \sigma$ represents the down quark.

We now turn our attention to the exotic fermions in Table~\ref{tab:5a}. By the same sequence of steps that led to the representation assignments in Table~\ref{tab:7}, we find that the $\U(2) \times \U(3)$ model interacting representation for the exotic fermions,
\begin{align*}
&\left(\textstyle\bigwedge^1 \overline{\eta} \otimes \bigwedge^0 \overline{\zeta} \otimes \sigma_{\mathnormal{L}} \right)  \oplus \left(\textstyle\bigwedge^1 \overline{\eta} \otimes \bigwedge^2 \overline{\zeta} \otimes \sigma_{\mathnormal{L}} \right) \oplus \left(\textstyle\bigwedge^0 \overline{\eta} \otimes \bigwedge^0 \overline{\zeta} \otimes \sigma_{\mathnormal{R}} \right) \\& \quad \oplus \left(\textstyle\bigwedge^2 \overline{\eta} \otimes \bigwedge^0 \overline{\zeta} \otimes \sigma_{\mathnormal{R}} \right) \oplus \left(\textstyle\bigwedge^0 \overline{\eta} \otimes \bigwedge^2 \overline{\zeta} \otimes \sigma_{\mathnormal{R}} \right) \oplus \left(\textstyle\bigwedge^2 \overline{\eta} \otimes \bigwedge^2 \overline{\zeta} \otimes \sigma_{\mathnormal{R}} \right),
\end{align*}
reduces to
\begin{align*}
&\left[ \left( \textstyle\bigwedge^3 \xi^{-} \oplus 1_{\U(3)_{\mathnormal{Q}}}\right)  \otimes \sigma_{\mathnormal{L}} \right] \oplus \left[ \left(\xi^{-} \oplus \xi^{+}\right)  \otimes \sigma_{\mathnormal{L}} \right] \oplus \sigma_{\mathnormal{R}}  \oplus \left(\textstyle\bigwedge^3 \xi^{-} \otimes \sigma_{\mathnormal{R}} \right) \oplus \left(\xi^{+}\otimes \sigma_{\mathnormal{R}} \right) \oplus \left(\xi^{-} \otimes \sigma_{\mathnormal{R}} \right) \\
& \quad = \sigma \oplus \left( \textstyle\bigwedge^3 \xi^{-} \otimes \sigma \right) \oplus \left( \xi^{+}\otimes \sigma \right) \oplus \left( \xi^{-}\otimes \sigma \right).
\end{align*}
Hence, the family of elementary fermions is extended to include exotic fermions: $\sigma$ represents an exotic electron-neutrino $N_E$, $\bigwedge^3 \xi^{-} \otimes \sigma$ represents an exotic electron $E$, $\xi^{+}\otimes \sigma$ represents an exotic up quark $U$, and $\xi^{-}\otimes \sigma$ represents an exotic down quark $D$. All these representations are listed in Table~\ref{tab:8}.
\begin{table}[ht]
\begin{tabular}{lcc}
\hline
\hline
Name                     & Symbol                 & Representation                              
	 \\
	 \hline 
	 \\
         Exotic neutrino      & $N_E$                 & $1_{\U(3)_{\mathnormal{Q}}}$ \\
         \\                                                               
         Exotic electron    & $E$                & $\bigwedge^3\xi^{-}$ \\
	 \\                                                               
          Exotic up quark   & $U$  &  $\xi^{+}$ \\
	 \\                                                               
         Exotic down quark & $D$  & $\xi^{-}$ \\ \\
	 \hline
	\end{tabular}
	\vspace{-5pt}
\caption{Exotic fermions as electrostrong $\U(3)_{\mathnormal{Q}}$ representations} \label{tab:8}
\end{table}

Next we shall establish the basic link between the electromagnetic group $\U(1)_{\mathnormal{Q}}$ and the electric charges of the electrostrong $\U(3)_{\mathnormal{Q}}$ representations. Towards that end, consider the representation $\lambda_{\mathnormal{Q}}^4$ of $\U(1)_{\mathnormal{Q}}$ on $\CC^1_4$. The same sort of arguments as those in Section~\ref{Sect:2.3} show that the representation $\xi_{\mathnormal{Q}}$ of $\U(1)_{\mathnormal{Q}} \times \SU(3)$ on $\CC^3$, which is obtained by composing the covering homomorphism $r_{\mathnormal{Q}} \colon \U(1)_{\mathnormal{Q}} \times \SU(3) \to \U(3)_{\mathnormal{Q}}$ with the representation $\xi^{+}$ of $\U(3)_{\mathnormal{Q}}$ on $\CC^3$, is equivalent to the representation $\lambda_{\mathnormal{Q}}^4 \otimes \rho$ of $\U(1)_{\mathnormal{Q}} \times \SU(3)$ on $\CC^1_4 \otimes \CC^3$. In a similar vein, if we consider the representation $\lambda_{\mathnormal{Q}}^{-2}$ of $\U(1)_{\mathnormal{Q}}$ on $\CC^1_{-2}$, then the representation $\xi_{\mathnormal{Q}}^{-}$ of $\U(1)_{\mathnormal{Q}} \times \SU(3)$ on $\CC^3$, which is obtained by composing $r_{\mathnormal{Q}}$ with the representation $\xi^{-}$ of $\U(3)_{\mathnormal{Q}}$ on $\CC^3$, is equivalent to the representation $\lambda_{\mathnormal{Q}}^{-2} \otimes \rho$ of $\U(1)_{\mathnormal{Q}} \times \SU(3)$ on $\CC^1_{-2} \otimes \CC^3$. Thus, every electrostrong $\U(3)_{\mathnormal{Q}}$ representation in Tables~\ref{tab:7} and \ref{tab:8} can be lifted to a representation of $\U(1)_{\mathnormal{Q}} \times \SU(3)$.  These representations are summarised in Table~\ref{tab:9}.
\begin{table}[ht]
\begin{tabular}{cc | cc}
\hline
\hline
 Symbol                 & Representation &  Symbol                 & Representation                              
	 \\
	 \hline 
	& & & \\
         $\nu_e$                 & $1_{\U(1)_{\mathnormal{Q}}} \otimes 1_{\SU(3)}$ &    $N_E$                 & $1_{\U(1)_{\mathnormal{Q}}} \otimes 1_{\SU(3)}$ \\
     & & &    \\                                                               
          $e$                & $\bigwedge^3(\lambda_{\mathnormal{Q}}^{-2} \otimes \rho)$ &   $E$                & $\bigwedge^3(\lambda_{\mathnormal{Q}}^{-2} \otimes \rho)$ \\
& & &	 \\                                                               
             $u$  &  $\lambda_{\mathnormal{Q}}^{4} \otimes \rho$ &   $U$  &  $\lambda_{\mathnormal{Q}}^{4} \otimes \rho$\\
	& & & \\                                                               
          $d$  & $\lambda_{\mathnormal{Q}}^{-2} \otimes \rho$ &   $D$  & $\lambda_{\mathnormal{Q}}^{-2} \otimes \rho$ \\ 
   & & &  \\
      \hline
	\end{tabular}
	\vspace{-5pt}
\caption{Elementary and exotic fermions as $\U(1)_{\mathnormal{Q}} \times \SU(3)$ representations} \label{tab:9}
\end{table}

To close this section, let us work out the electric charge values of the elementary and exotic fermions. Just as for the hypercharges analysed at the end of the previous section, the electric charge operator is an appropriate normalisation of the generator of the electromagnetic group $\U(1)_{\mathnormal{Q}}$. Inspection of Table~\ref{tab:9} indicates that every  fermion is assigned an electric charge, which is found to be a normalisation constant $q$ times the tensorial order of the corresponding $\U(1)_{\mathnormal{Q}}$ representation. In order for these charges to agree with the known assignments, we would have to pick $q=\frac{1}{6}$. The resulting electric charges for the different fermions are shown in Table~\ref{tab:10}. 
\begin{table}[ht]
\begin{tabular}{cc | cc}
\hline
\hline
 Symbol                 & Electric charge &  Symbol                 & Electric charge                             
	 \\
	 \hline
	& & & \\
         $\nu_e$                 & $\phantom{-}0$ &   $N_E$                 & $\phantom{-}0$  \\
     & & &    \\                                                               
          $e$                & $-1$ & $E$                & $-1$ \\
	& & & \\                                                               
             $u$  &  $\phantom{-}\frac{2}{3}$ &  $U$  &  $\phantom{-}\frac{2}{3}$ \\
 & & &	 \\                                                               
          $d$  & $-\frac{1}{3}$ &  $D$  & $-\frac{1}{3}$ \\ 
  & & &       \\                                                               
	 \hline
	\end{tabular}
	\vspace{-5pt}
\caption{Electric charge assignment for the elementary and exotic fermions} \label{tab:10}
\end{table}

\subsection{Anomaly cancellation}\label{Sect:2.5}
For non-abelian gauge theories, the elimination of chiral anomalies is one of the essential contraints new models of quarks and leptons must satisfy. In the standard model, the quarks and the leptons have chiralities that cancel the chiral anomaly. It is the purpose of this section to show that the cancellation of the chiral anomaly in the $\U(2) \times \U(3)$ model is achieved automatically.  

We begin by considering a gauge theory with gauge group $G$ in which we allow different matrix representations for the left- and right-handed fermions. According to the Adler-Bardeen theorem \cite{AB69} the theory is free of anomalies if and only if all the triangle diagram anomalies are absent. The condition for the cancelation is that
$$
\sum_{\text{reps.}} \tr \left\{ (\lambda_{i}^{\mathnormal{L}} \lambda_{j}^{\mathnormal{L}}+ \lambda_{j}^{\mathnormal{L}} \lambda_{i}^{\mathnormal{L}})\lambda_{k}^{\mathnormal{L}} \right\}  - \tr\left\{ (\lambda_{i}^{\mathnormal{R}} \lambda_{j}^{\mathnormal{R}}+ \lambda_{j}^{\mathnormal{R}} \lambda_{i}^{\mathnormal{R}})\lambda_{k}^{\mathnormal{R}} \right\} = 0,
$$
where the sum ranges over all the fermion representations (while the trace is taken over all the fermions in each representation) and where the Lie algebra matrices $\lambda_{i}^{\mathnormal{L}}$ and $\lambda_{i}^{\mathnormal{R}}$ are, respectively, the generators of the representation of $G$ on the left- and right-handed fermions. 

We can easily check that the anomaly cancellation condition is satisfied by the $\U(2) \times \U(3)$ model of Section~\ref{Sect:2.3}. For this purpose, we work out the explicit form for the $\U(2) \times \U(3)$ generators. The Lie algebra $\mathfrak{u}(2)$ of $\U(2)$ is the set of all $2 \times 2$ skew-adjoint matrices. Because $\U(1)_L \times \SU(2)$ covers $\U(2)$, $\U(1)_L \times \SU(2)$ and $\U(2)$ have isomorphic Lie algebras,
$$
\mathfrak{u}(2) \cong \mathfrak{u}(1)_L \oplus \mathfrak{su}(2).
$$
The four matrices $T_1=\frac{1}{2}i\sigma_1$, $T_2=\frac{1}{2}i\sigma_2$, $T_3=\frac{1}{2}i\sigma_3$ and $T_4=\frac{1}{2}i I_2$ provide a basis for $\mathfrak{u}(2) \cong \mathfrak{u}(1)_L \oplus \mathfrak{su}(2)$, where the Pauli matrices $\sigma_1$, $\sigma_2$ and $\sigma_3$ are explicitly given by
$$
\sigma_1=\left(\begin{array}{cc}
0 & 1 \\ 
1 & 0
\end{array} \right), 
\quad
\sigma_2=\left(\begin{array}{cc}
0 & -i \\ 
i & 0
\end{array} \right), 
\quad
\sigma_3=\left(\begin{array}{cc}
1 & 0 \\ 
0 & -1
\end{array} \right).
$$
Similarly, the Lie algebra $\mathfrak{u}(3)$ of $\U(3)$ is the set of all $3 \times 3$ skew-adjoint matrices. Since $\U(1)_C \times \SU(3)$ covers $\U(3)$, $\U(1)_C \times \SU(3)$ and $\U(3)$ have isomorphic Lie algebras,
$$
\mathfrak{u}(3) \cong \mathfrak{u}(1)_C \oplus \mathfrak{su}(3).
$$
The nine matrices $F_1 = \frac{1}{2}i\lambda_1$, $F_2 = \frac{1}{2}i\lambda_2,\dots, F_8 = \frac{1}{2}i\lambda_8$ and $F_9=\frac{1}{3} i I_3$ provide a basis for $\mathfrak{u}(3) \cong \mathfrak{u}(1)_C \oplus \mathfrak{su}(3)$, where the Gell-Mann matrices $\lambda_1,\lambda_2,\dots,\lambda_8$ are explicitly given by
\begin{align*}
\lambda_1 &=
\left(\begin{array}{ccc}
0 & 1 & 0 \\ 
1 & 0 & 0 \\ 
0 & 0 & 0
\end{array} \right),
\quad
\lambda_2=
\left(\begin{array}{ccc}
0 & -i & 0 \\ 
i & 0 & 0 \\ 
0 & 0 & 0
\end{array} \right),
\quad
\lambda_3=
\left(\begin{array}{ccc}
1 & 0 & 0 \\ 
0 & -1 & 0 \\ 
0 & 0 & 0
\end{array} \right), \\
\lambda_4 &=
\left(\begin{array}{ccc}
0 & 0 & 1 \\ 
0 & 0 & 0 \\ 
1 & 0 & 0
\end{array} \right),
\quad
\lambda_5=
\left(\begin{array}{ccc}
0 & 0 & -i \\ 
0 & 0 & 0 \\ 
i & 0 & 0
\end{array} \right), 
 \\
 \lambda_6 &=
\left(\begin{array}{ccc}
0 & 0 & 0 \\ 
0 & 0 & 1 \\ 
0 & 1 & 0
\end{array} \right),
\quad
\lambda_7=
\left(\begin{array}{ccc}
0 & 0 & 0 \\ 
0 & 0 & -i \\ 
0 & i & 0
\end{array} \right),
\quad 
\lambda_8=
{1\over\sqrt{3}}\left(\begin{array}{ccc}
1 & 0 & 0 \\ 
0 & 1 & 0 \\ 
0 & 0 & -2
\end{array} \right).
\end{align*}
A generic element of $\mathfrak{u}(2) \oplus \mathfrak{u}(3)$ is then
$$
\sum_{i=1}^4 \alpha_i T_i + \sum_{i=1}^9 \beta_i F_i,
$$
where the $\alpha_i$ and $\beta_i$ are arbitrary real constants. Herein, we  notice that the generator of the hypercharge group $\U(1)_Y$ may be taken to be $Y=T_4 - F_9$, while the generator of the complementary $\U(1)_Z$ to $\uS(\U(2) \times \U(3))$ may be taken as $Z=2 T_4 + 3 F_9$. If we set
\begin{align*}
\gamma_1&= \tfrac{1}{2}\left(\alpha_3 + \alpha_4\right), \\
\gamma_1&= -\tfrac{1}{2}\left(\alpha_3 - \alpha_4\right) ,\\
\delta_1 &= \tfrac{1}{2} \beta_3 + \tfrac{1}{2\sqrt{3}}\beta_8 + \tfrac{1}{3}\beta_9, \\
\delta_2 &= -\tfrac{1}{2} \beta_3 + \tfrac{1}{2\sqrt{3}}\beta_8 + \tfrac{1}{3}\beta_9 ,\\
\delta_3 &= - \tfrac{1}{\sqrt{3}}\beta_8 + \tfrac{1}{3}\beta_9,
\end{align*}
and regard the element $\sum_{i=1}^4 \alpha_i T_i + \sum_{i=1}^9 \beta_i F_i$ as a matrix acting on the left-handed multiplets, such a matrix is the direct sum of the following two blocks, 
\begin{gather*}
\left( \begin{array}{cccccccc}
i \left( \gamma_1+\delta_1+\delta_2+\delta_3 \right) & \frac{1}{2}\left( i \alpha_1 + \alpha_2 \right) & 0 & 0 & 0 & 0 & 0 & 0 \\
 \frac{1}{2}\left(i \alpha_1 - \alpha_2 \right) &  i  \left( \gamma_2+\delta_1+\delta_2+\delta_3 \right) & 0 & 0 & 0 & 0 & 0 & 0 \\
 0 & 0 &  i  \left( \gamma_1+\delta_1 \right) & \frac{1}{2}\left( i\beta_1 +  \beta_2 \right) &  \frac{1}{2}\left( i\beta_4 +  \beta_5 \right) &  \frac{1}{2}\left(i\alpha_1 + i \alpha_2 \right) & 0 & 0 \\
 0 & 0 & \frac{1}{2}\left(i\beta_1 -  \beta_2 \right) &  i \left( \gamma_1+\delta_2 \right) &  \frac{1}{2}\left( i\beta_6 - \beta_7 \right) & 0 &  \frac{1}{2}\left(i\alpha_1 + \alpha_2 \right) & 0 \\
 0 & 0 & \frac{1}{2}\left( i\beta_4 -  \beta_5 \right) & \frac{1}{2}\left(i \beta_6 - \beta_7 \right) &  i \left( \gamma_1+\delta_3 \right) & 0 & 0 &  \frac{1}{2}\left(i\alpha_1 + \alpha_2 \right) \\
 0 & 0 &  \frac{1}{2}\left(i\alpha_1 - \alpha_2 \right) & 0 & 0 & i \left( \gamma_2+\delta_1 \right) & \frac{1}{2}\left( i\beta_1 + \beta_2 \right) &  \frac{1}{2}\left( i\beta_4 + \beta_5 \right) \\
0 & 0 & 0 &  \frac{1}{2}\left(i\alpha_1 - \alpha_2 \right) & 0 &  \frac{1}{2}\left(i\beta_1 - \beta_2 \right) & i \left( \gamma_2+\delta_2 \right) &  \frac{1}{2}\left( i\beta_6 + \beta_7 \right) \\
0 & 0 & 0 & 0 &  \frac{1}{2}\left(i\alpha_1 - \alpha_2 \right) &  \frac{1}{2}\left( i\beta_4 - \beta_5 \right) & \frac{1}{2}\left(i\beta_6 - \beta_7 \right) & i \left( \gamma_2+\delta_3 \right)
 \end{array} \right)
\end{gather*}
and 
\begin{gather*}
\left( \begin{array}{cccccccc}
-i\gamma_1 & -\frac{1}{2}\left( i\alpha_1 - \alpha_2 \right) & 0 & 0 & 0 & 0 & 0 & 0 \\
 -\frac{1}{2}\left(i\alpha_1 + \alpha_2 \right) &  -i\gamma_2 & 0 & 0 & 0 & 0 & 0 & 0 \\
 0 & 0 & - i \left( \gamma_1+\delta_1+\delta_2 \right) & -\frac{1}{2}\left( i\beta_6 - \beta_7 \right) &  \frac{1}{2}\left( i\beta_4 - \beta_5 \right) &  -\frac{1}{2}\left(i\alpha_1 - \alpha_2 \right) & 0 & 0 \\
 0 & 0 & -\frac{1}{2}\left(i \beta_6 + \beta_7 \right) &  -i \left( \gamma_1+\delta_1+\delta_3 \right) &  -\frac{1}{2}\left( i\beta_1 - \beta_2 \right) & 0 &  -\frac{1}{2}\left(i\alpha_1 - \alpha_2 \right) & 0 \\
 0 & 0 & \frac{1}{2}\left( i\beta_4 + \beta_5 \right) & -\frac{1}{2}\left(i \beta_1 + \beta_2 \right) &  - i \left( \gamma_1+\delta_2+\delta_3 \right) & 0 & 0 &  -\frac{1}{2}\left(i\alpha_1 - \alpha_2 \right) \\
 0 & 0 &  -\frac{1}{2}\left(i\alpha_i + \alpha_2 \right) & 0 & 0 &  -i \left( \gamma_2+\delta_1+\delta_2 \right) & -\frac{1}{2}\left( i\beta_6 - \beta_7 \right) &  \frac{1}{2}\left( i\beta_4 - \beta_5 \right) \\
0 & 0 & 0 &  -\frac{1}{2}\left(i\alpha_1 + \alpha_2 \right) & 0 &  -\frac{1}{2}\left(i\beta_6 + \beta_7 \right) &  - i \left( \gamma_2+\delta_1+\delta_3 \right) & - \frac{1}{2}\left(i \beta_1 - \beta_2 \right) \\
0 & 0 & 0 & 0 &  -\frac{1}{2}\left(i\alpha_1 + \alpha_2 \right) &  \frac{1}{2}\left( i\beta_4 + \beta_5 \right) & -\frac{1}{2}\left(i\beta_1 + \beta_2 \right) &  - i \left( \gamma_2+\delta_2+\delta_3 \right)
 \end{array} \right).
\end{gather*}
Similarly, if we regard the element $\sum_{i=1}^4 \alpha_i T_i + \sum_{i=1}^9 \beta_i F_i$ as a matrix acting on the right-handed multiplets, such a matrix is the direct sum of the following two blocks,
\begin{gather*}
\left( \begin{array}{cccccccc}
i \left( \gamma_1+\gamma_2+\delta_1+\delta_2+\delta_3 \right) & 0 & 0 & 0 & 0 & 0 & 0 & 0 \\
0 & i \left(\delta_1+\delta_2+\delta_3 \right) & 0 & 0 & 0 & 0 & 0 & 0 \\
 0 & 0 &  i \left( \gamma_1+\gamma_2+\delta_1 \right) & \frac{1}{2}\left(i \beta_1 + \beta_2 \right) &  \frac{1}{2}\left( i\beta_4 +  \beta_5 \right) &  0 & 0 & 0 \\
 0 & 0 & \frac{1}{2}\left(i\beta_1 - \beta_2 \right) &   i \left( \gamma_1+\gamma_2+\delta_2 \right) &  \frac{1}{2}\left(i \beta_6 + \beta_7 \right) & 0 &  0 & 0 \\
 0 & 0 & \frac{1}{2}\left( i\beta_4 - \beta_5 \right) & \frac{1}{2}\left( i\beta_6 - \beta_7 \right) &   i \left( \gamma_1+\gamma_2+\delta_3 \right) & 0 & 0 & 0 \\
 0 & 0 & 0 & 0 & 0 &  i \delta_1 & \frac{1}{2}\left( i\beta_1 +  \beta_2 \right) &  \frac{1}{2}\left(i \beta_4 + \beta_5 \right) \\
0 & 0 & 0 & 0 & 0 &  \frac{1}{2}\left(i\beta_1 - \beta_2 \right) &  i\delta_2 &  \frac{1}{2}\left( i\beta_6 + \beta_7 \right) \\
0 & 0 & 0 & 0 &  0 &  \frac{1}{2}\left( i\beta_4 - \beta_5 \right) & \frac{1}{2}\left( i\beta_6 - \beta_7 \right) &  i\delta_3
 \end{array} \right)
\end{gather*}
and
\begin{gather*}
\left( \begin{array}{cccccccc}
0 & 0 & 0 & 0 & 0 & 0 & 0 & 0 \\
0 &  -i \left( \gamma_1+\gamma_2 \right) & 0 & 0 & 0 & 0 & 0 & 0 \\
 0 & 0 &  -i \left( \delta_1+\delta_2 \right) & -\frac{1}{2}\left( i\beta_6 - \beta_7 \right) &  \frac{1}{2}\left( i\beta_4 - \beta_5 \right) &  0 & 0 & 0 \\
 0 & 0 & -\frac{1}{2}\left(i\beta_7 + \beta_6 \right) &  -i \left( \delta_1+\delta_3 \right) &  -\frac{1}{2}\left( i\beta_1 - \beta_2 \right) & 0 &  0 & 0 \\
 0 & 0 & \frac{1}{2}\left( i\beta_4 +  \beta_5 \right) & -\frac{1}{2}\left(i \beta_1 + \beta_2 \right) &  -i \left( \delta_2+\delta_3 \right) & 0 & 0 & 0 \\
 0 & 0 & 0 & 0 & 0 &  -i \left(\gamma_1+\gamma_2+ \delta_1+\delta_2 \right) & -\frac{1}{2}\left( i\beta_6 - \beta_7 \right) &  \frac{1}{2}\left( i\beta_4 - \beta_5 \right) \\
0 & 0 & 0 & 0 & 0 &  -\frac{1}{2}\left(i\beta_6 +  \beta_7 \right) &   -i \left(\gamma_1+\gamma_2+ \delta_1+\delta_3 \right) &  -\frac{1}{2}\left( i\beta_1 - \beta_2 \right) \\
0 & 0 & 0 & 0 &  0 &  \frac{1}{2}\left( i\beta_4 + \beta_5 \right) & -\frac{1}{2}\left( i\beta_1 + \beta_2 \right) &   -i \left(\gamma_1+\gamma_2+ \delta_2+\delta_3 \right)
 \end{array} \right).
\end{gather*}
In this way, the representation matrices $T_i^{L,R}$ and $F_i^{L,R}$  can be systematically evaluated. For example, $T_i^{L}$ will correspond to the sum of the first two blocks with $\alpha_i=1$ and all the other constants set to zero, while $F_i^{R}$ will correspond to the sum of the two remaining blocks with $\beta_i=1$ and all the other constants set to zero.

Let us now consider the different possibilities for the cancellation of chiral anomalies. First, suppose, that $i$, $j$ and $k$ refer to the generators of $\mathfrak{su}(2)$. Then the condition is automatically satisfied because the result is proportional to $\tr (T_k^{L,R})=0$. Similarly, if $i$, $j$ and $k$ refer to the generators of $\mathfrak{su}(3)$, the condition is automatically satisfied because the result is proportional to $\tr (F_k^{L,R})=0$. On the other hand, if $i$ refers to a generator of $\mathfrak{su}(2)$ or $\mathfrak{su}(3)$ and $j$ and $k$ to $\mathfrak{u}(1)_Y$ or $\mathfrak{u}(1)_Z$, i.e., there is only a  $T_i$ or a $F_i$, then the condition is satisfied  because it contains $\tr (T_i^{L,R})=0$ or $\tr (F_i^{L,R})=0$. Suppose next that $i$ and $j$ refer to the $\mathfrak{su}(2)$ and $k$ to the $\mathfrak{u}(1)_Y$ or $\mathfrak{u}(1)_Z$. Then the required condition is proportional to $\tr(Y^L)=0$ or $\tr(Z^L)=0$ because the right part does not contribute. In contrast, if $i$ and $j$ refer to the $\mathfrak{su}(3)$ and $k$ to the $\mathfrak{u}(1)_Y$ or $\mathfrak{u}(1)_Z$, then the result is proportional to $\tr(Y^L)-\tr(Y^R)=0$ or $\tr(Z^L)-\tr(Z^R)=0$. On the other hand, if $i$ and $j$ refer to the $\mathfrak{u}(1)_Y$ and $k$ to the $\mathfrak{u}(1)_Z$, 
$$
\tr \{ (Y^L)^2 Z^L \} - \tr \{ (Y^R)^2 Z^R \} = 0
$$
and the condition is automatically satisfied. Similarly, if $i$ and $j$ refer to the $\mathfrak{u}(1)_Z$ and $k$ to the $\mathfrak{u}(1)_Y$, we have
$$
\tr \{ (Z^L)^2 Y^L \} - \tr \{ (Z^R)^2 Y^R \} = 0
$$
and so again the condition is satisfied. Finally, if $i$, $j$ and $k$ all refer to the generator of $\mathfrak{u}(1)_Y$ or $\mathfrak{u}(1)_Z$, we obtain the conditions
$$
\tr \{ (Y^L)^3 \} - \tr \{ (Y^R)^3 \} = 0
$$
or 
$$
\tr \{ (Z^L)^3 \} - \tr \{ (Z^R)^3 \} = 0.
$$
The cancellation of the chiral anomaly in the $\U(2) \times \U(3)$ model is thus ensured.

\section{Geometric formulation of the $\U(2)\times\U(3)$ model}\label{Sect:3}
It is now well known that gauge theories can be formulated as geometrical theories using the mathematical theory of connections on principal bundles; see, for instance, \cite{Trautman1979, DanielViallet80}. The purpose of this section is to give such a geometric description for the $\U(2) \times \U(3)$ model introduced above. All the notation introduced there remains valid. 

\subsection{Geometry of the $\U(2) \times \U(3)$ model}
We begin with an explicit enumeration of the basic geometric ingredients required to describe, at the classical level, a gauge theory with gauge group $\U(2) \times \U(3)$. Consider the following data:
\begin{enumerate}
\item A four-dimensional, oriented Lorentzian manifold $M$.
\item A principal bundle $P$ over $M$ with structure group $\U(2)\times \U(3)$.
\item A connection $\theta$ on $P$ with curvature $F_{\theta}$.
\item An equivariant map $\chi \colon P \to \bigwedge^2 \CC^2 \otimes \bigwedge^3 \CC^3$ corresponding to a section of the associated bundle $P \times_{\bigwedge^2 \eta \otimes \bigwedge^3 \zeta}\big(\bigwedge^2 \CC^2 \otimes \bigwedge^3 \CC^3 \big)$.
\item An equivariant map $\phi \colon P \to  \bigwedge^1\CC^2 \otimes \bigwedge^0 \CC^3$ corresponding to a section of the associated bundle $P \times_{\bigwedge^1\eta \otimes \bigwedge^0 \zeta}\big(\bigwedge^1\CC^2 \otimes \bigwedge^0 \CC^3 \big)$. 
\end{enumerate}
Typically, $M$ is taken to be the four-dimensional Minkowski space-time, in which case $P$ will be trivial since $M$ is a contractible topological space. The connection $\theta$ on $P$ may be written as $\theta = \omega + \gamma$, where $\omega$ and $\gamma$ are respectively $\mathfrak{u}(2)$-valued and  $\mathfrak{u}(3)$-valued one-forms on $P$. The associated curvature is simply $F_{\theta}= F_{\omega} + F_{\gamma}$. The equivariant map $\chi \colon P \to \bigwedge^2 \CC^2 \otimes \bigwedge^3 \CC^3$ is introduced in order to break down the symmetry from $\U(2) \times \U(3)$ to $\uS(\U(2) \times \U(3))$. Likewise, the equivariant map $\phi \colon P \to  \bigwedge^1\CC^2 \otimes \bigwedge^0 \CC^3$ is introduced with a view to breaking down the symmetry from $\uS(\U(2) \times \U(3))$ to the electrostrong group $\U(3)_Q$. Both $\chi$ and $\phi$ will be called \emph{Higgs fields}. 

When viewed in terms of this geometric set-up, each of the representations for the fermionic multiplets in Tables~\ref{tab:4} and \ref{tab:5a} corresponds to an interaction bundle which has that representation on its typical fibre. Given that these representations only use the standard representations $\eta$ of $\U(2)$ and $\zeta$ of $\U(3)$, as well as their conjugates $\overline{\eta}$ and $\overline{\zeta}$, one can obtain such an interaction bundle for each multiplet representation using only the vector bundles $P \times_{\eta} \CC^2$, $P \times_{\zeta} \CC^3$, $P \times_{\overline{\eta}} \CC^2$ and $P \times_{\overline{\zeta}} \CC^3$, as well as their exterior powers. However, in this section we shall ignore the fermionic multiplets and consider only the interactions between the gauge fields $\omega$ and $\gamma$ and the Higgs fields $\chi$ and $\phi$. At this point, we are primarily interested in the symmetry breaking in the gauge boson sector. The symmetry breaking in the fermionic sector will be treated elsewhere. 

A Lagrangian field theoretical model which can account for the spontaneous symmetry breaking of the gauge symmetry $\U(2) \times \U(3)$ can be given in terms of a Lagrangian differential $4$-form which describes the interaction between the gauge fields $\omega$ and $\gamma$ and the Higgs fields $\chi$ and $\phi$. Before writing it down explicitly, we must note that, since $\bigwedge^2 \CC^2 \otimes \bigwedge^3 \CC^3$ is linearly isomorphic to $\CC^1$ and $\bigwedge^1 \CC^2 \otimes \bigwedge^0 \CC^3$ is linearly isomorphic to $\CC^2$, the associated bundles $P \times_{\bigwedge^2\eta \otimes \bigwedge^3 \zeta}\big(\bigwedge^2\CC^2 \otimes \bigwedge^3 \CC^3 \big)$ and $P \times_{\bigwedge^1\eta \otimes \bigwedge^0 \zeta}\big(\bigwedge^1\CC^2 \otimes \bigwedge^0 \CC^3 \big)$ are both equipped with a hermitian metric. We denote by $\vert \chi \vert$ and $\vert \vert \phi \vert \vert$ the norms of $\chi$ and $\phi$ with respect to these hermitian metrics respectively. The Lagrangian model $4$-form is
$$
\mathcal{L} = -\frac{1}{4} \tr \left( F_{\omega} \wedge \ast F_{\omega} \right)-  \frac{1}{4} \tr \left( F_{\gamma} \wedge \ast F_{\gamma} \right) + \ast\vert \mathrm{d}^{\theta} \chi \vert^2 - \ast \frac{\lambda}{8}\left( \vert \chi \vert^2 - 1\right)^2 + \ast \vert \vert \mathrm{d}^{\theta} \phi \vert \vert^2 - \ast \frac{\mu}{8}\left( \vert\vert \phi \vert\vert^2 - 1\right)^2,
$$
where $\lambda$ and $\mu$ are non-negative constants, $\ast$ denotes the Hodge star operator determined by the metric on $M$, $\mathrm{d}^{\theta} \chi$ and $\mathrm{d}^{\theta} \phi$ are the covariant exterior derivatives of the Higgs fields $\chi$ and $\phi$ with respect to the connection $\theta$, and the norms arise from the metric on $M$ and the aforementioned hermitian metrics. Note that the first two terms are the typical Yang-Mills terms for the gauge fields $\omega$ and $\gamma$. 

The spontaneous symmetry breaking for the $\U(2) \times \U(3)$ model can now be described as follows (cf.~\cite{KKL86,CDNR87}). The Higgs field $\chi$, whose self-interaction is the potential $4$-form $\ast \tfrac{\lambda}{8}\left( \vert \chi \vert^2 - 1\right)^2$, breaks down the gauge symmetry from $\U(2) \times \U(3)$ to $\uS(\U(2) \times \U(3))$.  Similarly, the Higgs field $\phi$, whose self-interaction is the potential $4$-form $\ast \tfrac{\mu}{8}\left( \vert\vert \phi \vert\vert^2 - 1\right)^2$, breaks down the gauge symmetry from $\uS(\U(2) \times \U(3))$ to $\U(3)_Q$. The next two sections explain the breakdowns in detail.

\subsection{Breakdown of $\U(2) \times \U(3)$ to $\uS(\U(2) \times \U(3))$}
In this section we shall show how to use the Higgs field $\chi$ to choose an appropriate $\uS(\U(2) \times \U(3))$ subbundle of $P$ to which $\theta$ will reduce. The component of the reduced connection $\theta$ will then play the role of the standard model gauge field. 

Note that the minimum of the potential function $U(\xi)=\frac{\lambda}{8}\left( \vert \xi \vert^2 - 1\right)^2$, defined on $\CC^1 \cong \bigwedge^2 \CC^2 \otimes \bigwedge^3 \CC^3$, is attained at any vector $\xi$ with length $1$. Let us fix one such vector $\xi_0$. With respect to this vector, we can always perform a $\U(2) \times \U(3)$ gauge transformation to a unitary gauge in which $\chi$ is a symmetry-breaking Higgs field on $P$, i.e., it maps all of $P$ onto the orbit of $\U(2) \times \U(3)$ through $\xi_0 \in \CC^1 \cong \bigwedge^2 \CC^2 \otimes \bigwedge^3 \CC^3$. Consequently, $\chi^{-1}(\xi_0)$ is a subbundle of $P$ which we denote by $P_{\xi_0}$. Its structure group is the isotropy subgroup of $\xi_0$ in $\U(2) \times \U(3)$. If $(u,v)$ is a typical element of $\U(2) \times \U(3)$ we immediately see that $(u,v)$ fixes $\xi_0$ if and only if 
$$
(\det u)(\det v)=1.
$$
Thus the subbundle defined by $P_{\xi_0}$ has structure group $\uS(\U(2) \times \U(3))$.

Let us examine what happens to the connection $\theta$. Consider the basis $\{T_1,T_2,T_3,T_4\}$ of $\mathfrak{u}(2) \cong \mathfrak{u}(1)_L \oplus \mathfrak{su}(2)$ and the basis $\{F_1,F_2,\dots,F_9\}$ of $\mathfrak{u}(3) \cong \mathfrak{u}(1)_C \oplus \mathfrak{su}(3)$ introduced in Section~\ref{Sect:2.5}, put $T_L = T_4$ and $F_C = F_9$, and recall that $\mathfrak{u}(1)_Y$ is generated by $Y=T_L - F_C$, whilst $\mathfrak{u}(1)_Z$ is generated by $Z=2T_L + 3F_C$. Since $\theta$ is $\mathfrak{u}(2) \oplus \mathfrak{u}(3)$-valued, we may write
$$
\theta = 
\left(\sum_{i=1}^3 \omega^{i} g' T_i   + \omega^{L} g'_L T_L\right)
+
\left(\sum_{i=1}^8 \gamma^{i} g'' F_i  + \gamma^C g''_C T_C\right),
$$
where $g'$, $g''$, $g'_L$ and $g''_C$ are coupling constants to be determined by empirical data. Notice that the first sum corresponds to the gauge field $\omega$ and the second to the gauge field $\gamma$, and that there is one coupling constant for each factor of $\mathfrak{u}(2) \oplus \mathfrak{u}(3)$. Now, as discussed in Section~\ref{Sect:2.2}, $\U(1)_Y \times \SU(2) \times \SU(3)$ covers $\uS(\U(2)\times \U(3))$, so that both groups have isomorphic Lie algebras,
$$
\mathfrak{s}(\U(2) \times \U(3)) \cong \mathfrak{u}(1)_Y \oplus \mathfrak{su}(2) \oplus \mathfrak{su}(3).
$$
Moreover, if $\mathfrak{m}$ is the subspace of $\mathfrak{u}(2) \oplus \mathfrak{u}(3)$ generated by $a T_L + b F_C$ for some constants $a$ and $b$ such that $\frac{1}{2}a + \frac{1}{3} b \neq 0$ then $\mathrm{Ad}_h (\mathfrak{m}) \subset \mathfrak{m}$ for every $h \in \uS(\U(2) \times \U(3))$. Consequently, if $\theta_{\mathfrak{s}(\U(2)\times\U(3))}$ and $\theta_{\mathfrak{m}}$ are the components of $\theta$ relative to the decomposition
$$
\mathfrak{u}(2) \oplus \mathfrak{u}(3) =\left[\mathfrak{s}(\U(2) \times \U(3)) \right]\oplus \mathfrak{m} \cong \left[\mathfrak{u}(1)_Y \oplus \mathfrak{su}(2) \oplus \mathfrak{su}(3) \right] \oplus \mathfrak{m},
$$
then $\theta_{\mathfrak{s}(\U(2)\times\U(3))} \vert TP_{\xi_0}$ is a connection on $P_{\xi_0}$ and $\theta_{\mathfrak{m}} \vert TP_{\xi_0}$ is a tensorial $1$-form on $P_{\xi_0}$ with values in $\mathfrak{m}$. 

We claim that after the symmetry group $\U(2) \times \U(3)$ has been reduced to $\uS(\U(2) \times \U(3))$, the interaction term in the Lagrangian model $4$-form can be used to impart mass to the tensorial component $\theta_{\mathfrak{m}} \vert TP_{\xi_0}$. In order to substantiate the claim, we first observe that
$$
\theta \bigcdot \xi_0 = \omega^{L} g'_L T_L \bigcdot \xi_0 + \gamma^{9} g''_C F_C \bigcdot \xi_0= i\left(\omega^{L} g'_L  \xi_0 + \gamma^{C} g''_C  \xi_0 \right).
$$
The relevant term in the Lagrangian $4$-form is then determined by the bilinear form
$$
\vert \mathrm{d}^{\theta} \xi_0 \vert^2 = \vert \theta \bigcdot \xi_0 \vert^2 = g'^2_L (\omega^L)^2 + g''^2_C (\gamma^C)^2 + 2 g'_L g''_C \omega^L \gamma^C. 
$$
We wish to write this bilinear form in terms of two new bases $\{ \overline{T}_1,  \overline{T}_2,  \overline{T}_3,  \overline{T}_L \}$ and $\{ \overline{F}_1,\overline{F}_2,\dots,\overline{F}_C \}$ such that it becomes diagonal. It will suffice to find an orthogonal matrix
$$
A = \left( \begin{array}{cc} \cos \varphi & \sin \varphi \\ -\sin \varphi & \cos \varphi \end{array} \right)
$$  
for some $\varphi$ so that $\overline{T}_i = g' T_i$ for $i=1,2,3$, $\overline{F}_i = g'' F_i$ for $i=1,2,\dots,8$, and
\begin{align*}
\overline{T}_L &= (\cos \varphi) g'_L T_L + (-\sin \varphi) g''_C F_C,\\
\overline{F}_C &= (\sin \varphi) g'_C T_C + (\cos \varphi) g''_C F_C.
\end{align*}
In term of the new bases,
$$
\theta = \sum_{i=1}^3 \overline{\omega}^{i} \overline{T}_i + \sum_{i=1}^8 \overline{\gamma}^{i} \overline{F}_i +  \overline{\omega}^{L} \overline{T}_L + \overline{\gamma}^{C} \overline{F}_C
$$
where $\overline{\omega}^{i}= \omega^{i}$ for $i=1,2,3$, $\overline{\gamma}^{i}=\gamma^{i}$ for $i=1,2,\dots,8$, and
\begin{align*}
\omega^L &= ( \cos \varphi)\overline{\omega}^L + (\sin \varphi)\overline{\gamma}^C ,\\
\gamma^C &=  (-\sin \varphi)\overline{\omega}^L  + (\cos \varphi)\overline{\gamma}^C .
\end{align*}
By a straightforward calculation
\begin{align*}
\vert \mathrm{d}^{\theta} \xi_0 \vert^2 =& \left[ g'^2_L \cos^2 \varphi + g''^2_C \sin^2 \varphi - 2 g'_L g''_C \sin \varphi \cos \varphi \right] (\overline{\omega}^L)^2  \\
&+\left[ g'^2_L \sin^2 \varphi + g''^2_C \cos^2 \varphi + 2 g'_L g''_C \sin \varphi \cos\varphi \right] (\overline{\gamma}^C)^2 \\
&+ \left[ 2 g'^2_L \sin \varphi \cos \varphi - 2 g''^2_C \sin \varphi \cos \varphi + 2 g'_L g''_C \cos^2\varphi - 2 g'_L g''_C \sin^2\varphi \right] \overline{\omega}^L \overline{\gamma}^C.
\end{align*}
We wish to choose $\varphi$ so that $\vert \mathrm{d}^{\theta} \xi_0 \vert^2$ will be diagonalised. We also want one of the fields to be a $\uS(\U(2) \times \U(3))$ gauge field and hence a connection on the surviving $\uS(\U(2) \times \U(3))$ bundle. Thus it should have values in the Lie algebra generated by $\{ T_1, T_2, T_3,Y \}$ and $\{F_1,F_2,\dots,F_8\}$ since these are the generators of $\mathfrak{s}(\U(2)\times\U(3))$ which annihilate the vector $\xi_0$ we used to compute $\vert \mathrm{d}^{\theta} \xi_0 \vert^2$ above. We choose to let $\overline{\omega}^4$ be the component of the surviving connection on $P_{\xi_0}$ so we want $\overline{T}_L= c Y$ for some constant $c$. Thus, $(\cos \varphi) g'_L= c$ and $(\sin \varphi) g''_C=c$. Therefore 
$$
\tan \varphi = \frac{g'_L}{g''_C}, \quad \sin \varphi = \frac{g'_L}{\sqrt{g'^2_L+g''^2_C}}, \quad \cos \varphi = \frac{g''_C}{\sqrt{g'^2_L+g''^2_C}}.
$$
From this we deduce that
$$
\vert \mathrm{d}^{\theta} \xi_0 \vert^2 = \frac{4 (g'_L g''_C)^2}{g'^2_L + g''^2_C} (\overline{\gamma}^C)^2.
$$
This last equation shows that, if we take $a=g'^2_L/\sqrt{g'^2_L + g''^2_C}$ and $b=g''^2_C/\sqrt{g'^2_L + g''^2_C}$ in the definition of $\mathfrak{m}$, then $\theta_{\mathfrak{m}}\vert TP_{\xi_0}=(\overline{\gamma}^C \vert TP_{\xi_0}) \overline{F}_C$ represents a massive gauge field with mass
$$
\frac{2 g'_L g''_C}{\sqrt{g'^2_L + g''^2_C}}.
$$
We see also that
$$
\theta_{\mathfrak{s}(\U(2)\times\U(3))} \vert TP_{\xi_0}=(\overline{\omega}^L \vert T P_{\xi_0}) \overline{T}_L + \sum_{i=1}^3 (\overline{\omega}^i \vert T P_{\xi_0})  \overline{T}_i + \sum_{i=1}^8 (\overline{\gamma}^i \vert T P_{\xi_0})  \overline{F}_i
$$
represents a massless gauge field. Since
$$
\overline{T}_L = \frac{g'_L g''_C}{\sqrt{g'^2_L + g''^2_C}}Y
$$
it follows that $g'_L g''_C / \sqrt{g'^2_L + g''^2_C}$ plays the role of the coupling constant of the $\U(1)_Y$ part of the spontaneously broken gauge group $\uS(\U(2) \times \U(3))$. Moreover, as a consequence of the definition, we obtain
\begin{align*}
\overline{\omega}^L &= \frac{g''_C \omega^L - g'_L \gamma^C}{\sqrt{g'^2_L + g''^2_C}}, \\
\overline{\gamma}^C &= \frac{g'_L \omega^L + g''_C \gamma^C}{\sqrt{g'^2_L + g''^2_C}}.
\end{align*}

Now according to Sections~\ref{Sect:2.3} and \ref{Sect:2.5}, we must also uphold the fact that the fundamental and exotic fermions of the model are charged under the complementary $\U(1)_Z$ to $\uS(\U(2) \times \U(3))$. This means that the subspace $\mathfrak{m}$ must be taken to be equal to $\mathfrak{u}(1)_Z$ and, therefore, we shall write $\theta_Z$ instead of $\theta_{\mathfrak{m}}$. The condition $\overline{F}_C= Z$ now implies that $g'^2_L/\sqrt{g'^2_L + g''^2_C}=2$ and $g''^2_C/\sqrt{g'^2_L + g''^2_C}=3$. Thus, the coupling constant $g''_C$ of the $\mathfrak{u}(1)_C$ factor is determined by the coupling constant $g'_L$ of the $\mathfrak{u}(1)_L$ factor via the relation
$$
g''_C = \sqrt{\frac{3}{2}}g'_L.
$$
Consequently, if we simply put $g= \sqrt{\frac{3}{5}} g'_L$, as we shall do in the sequel, $\theta_Z \vert TP_{\xi_0}= (\overline{\gamma}^C \vert TP_{\xi_0}) Z$ is a massive gauge field whose mass is $2 g$.  The corresponding value for the coupling constant of the $\U(1)_Y$ factor of $\uS(\U(2) \times \U(3))$ will be $g$. Finally, the two expressions above for $\overline{\omega}^L$ and $\overline{\gamma}^C$ reduce to
\begin{align*}
\overline{\omega}^L &=\sqrt{\frac{3}{5}}\omega^L - \sqrt{\frac{2}{5}}\gamma^C, \\
\overline{\gamma}^C &= \sqrt{\frac{2}{5}}\omega^L + \sqrt{\frac{3}{5}}\gamma^C.
\end{align*}

\subsection{Breakdown of $\uS(\U(2) \times \U(3))$ to $\U(3)_Q$}
Our task now is to show how to use the Higgs field $\phi$ to choose a candidate for an appropriate $\U(3)_Q$ subbundle of $P_{\xi_0}$ on which $\theta_{\mathfrak{s}(\U(2)\times\U(3))} \vert TP_{\xi_0}$ reduces. As we shall see, the components of the reduced connection will play the role of the gauge fields which occur in the usual standard model. 

As before, we begin by noticing that the minimum of the potential function $V(\varrho)=\frac{\mu}{8}(\vert\vert \varrho \vert\vert^2 - 1)^2$, defined on $\CC^2 \cong \bigwedge^1\CC^2 \otimes \bigwedge^0 \CC^3$, is attained at any vector $\varrho$ with length $1$. We shall choose $\varrho = \binom{0}{1}$ for our analysis. On the other hand, it is not hard to verify that the restriction of the Higgs field $\phi\vert P_{\xi_0} \colon P_{\xi_0} \to \bigwedge^1\CC^2 \otimes \bigwedge^0 \CC^3$ is equivariant with respect to the action of $\uS(\U(2) \times \U(3))$, so a Higgs field too. And just as in the previous section, we can perform a $\uS(\U(2) \times \U(3))$ gauge transformation to ensure that $\phi\vert P_{\xi_0}$ is a symmetry-breaking Higgs field on $P_{\xi_0}$, i.e., it maps all of $P_{\xi_0}$ onto the orbit of $\uS(\U(2) \times \U(3))$ through $\binom{0}{1} \in \CC^2\cong \bigwedge^1\CC^2 \otimes \bigwedge^0 \CC^3$. Thus, $(\phi\vert P_{\xi_0})^{-1}\binom{0}{1}$ is a subbundle of $P_{\xi_0}$ which we denote by $Q_{\xi_0}$. Its structure group is the isotropy group of $\binom{0}{1}$ in $\uS(\U(2) \times \U(3))$. If $(u,v)$ is a typical element of $\uS(\U(2) \times \U(3))$ one can easily check that $\left[\left( \bigwedge^1 \eta \otimes \bigwedge^0 \zeta\right)(u,v)\right]\binom{0}{1}=\binom{0}{1}$ if and only if
$$
u = \left( \begin{array}{cc}(\det v)^{-1} & 0 \\ 0 & 1 \end{array}\right).
$$
In particular, an element of the form $(u,(\det v)^{-1} v)$ is in the isotropy group of $\binom{0}{1}$ in $\uS(\U(2)\times \U(3))$ if and only if
$$
u = \left( \begin{array}{cc}(\det v)^{2} & 0 \\ 0 & 1 \end{array}\right).
$$
Therefore the subbundle defined by $Q_{\xi_0}$ has structure group $\U(3)_Q$. 

The following step is to delve into what happens to the connection $\theta_{\mathfrak{s}(\U(2)\times\U(3))} \vert TP_{\xi_0}$. Let us first, however, lighten up the notation by setting $\theta_{\xi_0}=\theta_{\mathfrak{s}(\U(2)\times\U(3))} \vert TP_{\xi_0}$, $\beta_{\xi_0} = \overline{\omega}^L\vert TP_{\xi_0}$, $\omega_{\xi_0}^i= \overline{\omega}^i\vert TP_{\xi_0}=\omega^i\vert TP_{\xi_0}$ for $i=1,2,3$, and $\gamma_{\xi_0}^i=\overline{\gamma}^i\vert TP_{\xi_0}=\gamma^i\vert TP_{\xi_0}$ for $i=1,2,\dots,8$. Thus, we may write
$$
\theta_{\xi_0} = \beta_{\xi_0} g Y + \sum_{i=1}^3 \omega_{\xi_0}^{i} g' T_i + \sum_{i=1}^8 \gamma_{\xi_0}^{i} g'' F_i.
$$
By appealing to Section~\ref{Sect:2.2}, we know that $\U(1)_Q \times \SU(3)$ covers $\U(3)_Q$ and therefore both groups have isomorphic Lie algebras,
$$
\mathfrak{u}(3)_Q \cong \mathfrak{u}(1)_Q \oplus \mathfrak{su}(3).
$$
Observe that $\mathfrak{u}(1)_Q$ is generated by $T_3 + Y$ since this is the generator of $\mathfrak{u}(3)_Q$ which annihilates the explicit vector $\binom{0}{1}$. Observe also that if $\mathfrak{m}$ is the subspace of $\mathfrak{s}(\U(2)\times\U(3))$ generated by $\{T_1,T_2, a T_3 + b Y\}$ and $\{F_1,F_2,\dots,F_8\}$ for some constants $a$ and $b$ such that $a T_3 + b Y \neq 0$ then $\mathrm{Ad}_h(\mathfrak{m})\subset \mathfrak{m}$ for every $h \in \U(3)_Q$. If $\theta_{\xi_0,\mathfrak{u}(3)_Q}$ and $\theta_{\xi_0,\mathfrak{m}}$ are the components of $\theta_{\xi_0}$ relative to the decomposition
$$
\mathfrak{s}(\U(2) \times \U(3)) \cong \mathfrak{u}(3)_Q \oplus \mathfrak{m},
$$
then $\theta_{\xi_0,\mathfrak{u}(3)_Q}\vert TQ_{\xi_0}$ is a connection on $Q_{\xi_0}$ and $\theta_{\xi_0,\mathfrak{m}}\vert T Q_{\xi_0}$ is a tensorial $1$-form on $Q_{\xi_0}$ with values in $\mathfrak{m}$.

We claim that the three components of $\theta_{\xi_0,\mathfrak{m}}\vert TQ_{\xi_0}$ will represent massive gauge fields. Note that the generators $F_1,\dots,F_8$ of $\mathfrak{su}(3)$ act trivially on the vector $\binom{0}{1}$ and hence
$$
\theta_{\xi_0} \bigcdot\! \left(\begin{array}{c} 0 \\ 1\end{array}\right)= \beta_{\xi_0} g \,Y\bigcdot\! \left(\begin{array}{c} 0 \\ 1\end{array}\right) + \sum_{i=1}^3 \omega_{\xi_0}^{i} g' \,T_i \bigcdot \!\left(\begin{array}{c} 0 \\ 1\end{array}\right)=\left(\begin{array}{c} \frac{1}{2}i g' \omega_{\xi_0}^1 +  \frac{1}{2}g'\omega_{\xi_0}^2 \\ \frac{1}{2}i g\beta_{\xi_0} - \frac{1}{2}i g' \omega_{\xi_0}^3 \end{array}\right).
$$
The Lagrangian $4$-form then yields a mass-producing term dictated by the bilinear form 
$$
\left\vert\left\vert \theta_{\xi_0} \bigcdot\! \left(\begin{array}{c} 0 \\ 1\end{array}\right) \right\vert\right\vert^2 = \frac{1}{4} g^2 (\omega_{\xi_0}^1)^2 + \frac{1}{4} g^2 (\omega_{\xi_0}^2)^2 + \frac{1}{4} g^2 (\omega_{\xi_0}^3)^2 + \frac{1}{4} g'^2 (\beta_{\xi_0})^2 - \frac{1}{2} g g' \omega_{\xi_0}^3 \beta_{\xi_0}.
$$
In order to identify the components to be assigned masses we shall write this bilinear form in terms of two new bases $\{\overline{T}_1,\overline{T}_2,\overline{T}_3,\overline{Y}\}$ and $\{\overline{F}_1,\overline{F}_2,\dots, \overline{F}_8\}$ such that it becomes diagonal. Since the first two terms are already in diagonal form it will suffice to choose an orthogonal matrix
$$
A = \left( \begin{array}{cc} \cos \psi & \sin \psi \\ -\sin \psi & \cos \psi \end{array} \right)
$$  
for some $\psi$ so that $\overline{T}_1 = g T_1$, $\overline{T}_2 = g T_2$, $\overline{F}_i = g'' F_i$ for $i=1,2,\dots,8$, and
\begin{align*}
\overline{T}_3 &= (\cos \psi) g T_3 + (-\sin \psi) g' Y, \\
\overline{Y} &= (\sin \psi) g T_3 + (\cos \psi) g' Y.
\end{align*}
Relative to these bases, 
$$
\theta_{\xi_0}= \overline{\beta}_{\xi_0} \overline{Y} + \sum_{i=1}^3 \overline{\omega}_{\xi_0}^{i} \overline{T}_i + \sum_{i=1}^8 \overline{\gamma}_{\xi_0}^{i} \overline{F}_i
$$
where $\overline{\omega}_{\xi_0}^{1}=\omega_{\xi_0}^{1}$, $\overline{\omega}_{\xi_0}^{2}=\omega_{\xi_0}^{2}$, $\overline{\gamma}_{\xi_0}^{i}=\gamma_{\xi_0}^{i}$ for $i=1,2,\dots,8$, and
\begin{align*}
\omega_{\xi_0}^{3} &= (\cos \psi) \overline{\omega}_{\xi_0}^{3} + (\sin \psi) \overline{\beta}_{\xi_0}, \\
\beta_{\xi_0} &= (-\sin \psi) \overline{\omega}_{\xi_0}^{3} + (\cos \psi) \overline{\beta}_{\xi_0}.
\end{align*}
A simple calculation leads to
\begin{align*}
\left\vert\left\vert \theta_{\xi_0} \bigcdot\! \left(\begin{array}{c} 0 \\ 1\end{array}\right) \right\vert\right\vert^2 =& \frac{1}{4}g^2 (\overline{\omega}_{\xi_0}^1)^2 + \frac{1}{4}g^2 (\overline{\omega}_{\xi_0}^2)^2 +  \frac{1}{4}\left( g \cos \psi + g' \sin \psi \right)^2 (\overline{\omega}_{\xi_0}^3)^2  \\
&+  \frac{1}{4} \left( g \sin \psi -  g' \cos \psi \right)^2 (\overline{\beta}_{\xi_0})^2 \\
& +  \frac{1}{4} \left[ \left( g^2 - g'^2 \right) \sin\psi \cos\psi - g g' (\cos^2 \psi - \sin^2 \psi) \right] \overline{\omega}_{\xi_0}^3\overline{\beta}_{\xi_0}.
\end{align*}
We wish to choose $\psi$ in such a way that $\left\vert\left\vert \theta_{\xi_0} \bigcdot \binom{0}{1} \right\vert\right\vert^2$ will be diagonalised. We also want the field $\overline{\beta}_{\xi_0}$ to be one of the components of the surviving connection on $Q_{\xi_0}$ with values in the Lie algebra $\mathfrak{u}(1)_Q$. This, in turn, means that we must impose the condition $\overline{Y}=c(T_3 + Y)$ for some constant $c$. It follows that $(\sin\psi) g=c$ and $(\cos \psi) g'=c$. Thus
$$
\sin \psi = \frac{g'}{\sqrt{g^2 + g'^2}}, \quad \cos \psi =\frac{g}{\sqrt{g^2 + g'^2}}. 
$$
Replacing these equalities into the expression for  $\left\vert\left\vert \theta_{\xi_0} \bigcdot \binom{0}{1} \right\vert\right\vert^2$ we obtain
$$
\left\vert\left\vert \theta_{\xi_0} \bigcdot\! \left(\begin{array}{c} 0 \\ 1\end{array}\right) \right\vert\right\vert^2 = \frac{1}{4}g^2 (\overline{\omega}_{\xi_0}^1)^2 + \frac{1}{4}g^2 (\overline{\omega}_{\xi_0}^2)^2 + \frac{1}{4}\sqrt{g^2 + g'^2}(\overline{\omega}_{\xi_0}^3)^2.
$$
Hence $\overline{\omega}_{\xi_0}^1$, $\overline{\omega}_{\xi_0}^2$ and $\overline{\omega}_{\xi_0}^3$ represent massive gauge fields with masses $\frac{1}{2} g$, $\frac{1}{2} g$ and $\frac{1}{2} \sqrt{g^2 + g'^2}$, respectively. We see also that $(\overline{\beta}_{\xi_0}\vert TQ_{\xi_0}) \overline{Y}$ represents a massless gauge field. Since $
\overline{Y} = \frac{gg'}{\sqrt{g^2+g'^2}}(T_3 + Y)$, its eigenvalues are multiples of $\frac{gg'}{\sqrt{g^2+g'^2}}$ so that this gauge field interacts with charged particles whose fundamental unit of charge is $\frac{gg'}{\sqrt{g^2+g'^2}}$. Furthermore, as is evident from the above expressions, we conclude that
\begin{align*}
\overline{\omega}_{\xi_0}^3 &= \frac{g \omega_{\xi_0}^3 - g'\beta_{\xi_0}}{\sqrt{g^2+g'^2}}, \\
\overline{\beta}_{\xi_0} &= \frac{g' \omega_{\xi_0}^3 + g \beta_{\xi_0}}{\sqrt{g^2+g'^2}}.
\end{align*}
If we identify $\psi$ with the Weinberg angle then we may identify $\overline{\omega}_{\xi_0}^3$ with the massive neutral boson field $Z^0$ and $\overline{\beta}_{\xi_0}$ with the massless electromagnetic vector potential $A$ in the traditional form of the standard model. We would also like to emphasise the fact that the fields $\overline{\omega}_{\xi_0}^1$ and $\overline{\omega}_{\xi_0}^2$ represent charged particles as they experience mixing under the action of the electromagnetic group $\U(1)_Q$. Thus we may identify the complex fields $\overline{\omega}_{\xi_0}^1 \pm i \overline{\omega}_{\xi_0}^2$ with the massive charged weak bosons $W^{\pm}$ of the standard model.

\section{Conclusion}\label{Sect:4}
In this paper, we have presented a $\U(2)\times \U(3)$ gauge extension of the standard model based on the premise that the confinement principle dictates 
\begin{itemize}
\item $1_{\SU(3)}$ and $\bigwedge^3\rho$ must be distinguished; 
\item the standard model's interacting-particle representation must be rewritten accordingly;
\item the hypercharge must be distributed.
\end{itemize}
These considerations implied the existence of a finite-to-one mapping from $\U(1)\times \SU(2)\times \SU(3)$ into $\U(2)\times \U(3)$ which was then used to build the model. Once we found a regular pattern of $\U(2)\times \U(3)$ representations for the fundamental fermions, it was easy to postulate many other similar representations with various combinations of chiralities as candidates for exotic particles, given the necessity of cancelling out new chiral anomalies arising from the extra degree of freedom inherent in the gauge group $\U(2)\times \U(3)$. We have analysed the properties of what we believe, at this point, is the most natural candidate for exotic fermions and proved that the requirements for anomaly cancellation are satisfied. We showed, additionally, that when the $\U(2) \times \U(3)$ symmetry is spontaneously broken down to the electrostrong group $\U(3)_Q$, the $\U(2) \times \U(3)$ representations for the fermionic multiplets decompose into direct sums of representations of $\U(3)_Q$. By lifting these $\U(3)_Q$ representations to representations of $\U(1)_Q \times \SU(3)$, we were able to determine the electric charges carried by the fermionic multiplets. 

We have also discussed a geometric formulation of the $\U(2) \times \U(3)$ model in the gauge boson sector and the symmetry breaking process. We found that the symmetry breaking of the $\U(2) \times \U(3)$ gauge group must be carried out in two steps: from $\U(2) \times \U(3)$ to the standard model group $\uS(\U(2) \times \U(3))$, and  from there to the electrostrong group $\U(3)_Q$. 

The numerous possibilities of exotic fermions suggested by the model pose many new questions besides the exploration of the dynamics, the symmetry breaking process in the fermonic sector and the phenomenological implications. We hope to pursue these topics in the near future.


\addcontentsline{toc}{section}{Bibliography}

\end{document}